\documentclass[reprint,amsmath,amssymb,aps,longbibliography]{revtex4-1}
\usepackage{graphicx}% Include figure files
\graphicspath{{images/}}
\usepackage{dcolumn}% Align table columns on decimal point
\usepackage{bm}% bold math

\usepackage{ifpdf}
\usepackage{lineno,amsfonts}
\usepackage{bbm}
\usepackage{mathrsfs}
\usepackage{upgreek}
\usepackage{mathtools}
\usepackage{epstopdf} 
\usepackage{setspace}
\usepackage{hyperref}
\usepackage{float}
\usepackage{natbib}
\usepackage{ifpdf}
\usepackage{qcircuit}
\usepackage[usenames,dvipsnames]{xcolor}
\usepackage{graphicx}
\usepackage{cellspace}
\usepackage{multirow}
\usepackage{cleveref}
\usepackage{enumitem}

\allowdisplaybreaks

\newcommand{\minus}{\texttt{-}}
\usepackage{amsfonts}% http://ctan.org/pkg/amsfonts
\let\oldequiv\equiv
\renewcommand{\equiv}[1][0pt]{\mathrel{\raisebox{#1}{$\oldequiv$}}}
\newcommand{\ket}[1]{\vert{#1}\rangle}
\newcommand{\bra}[1]{\langle{#1}\vert}

\begin{document}

\title{Exploration of an augmented set of Leggett-Garg inequalities using a non-invasive continuous in time velocity measurement}
\author{Shayan Majidy} 
\email{smajidy@uwaterloo.ca}
\author{Hemant Katiyar} 
\author{Galit Anikeeva}
\affiliation{Institute for Quantum Computing and Department of Physics and Astronomy, University of Waterloo, Waterloo, Ontario N2L 3G1, Canada} 
\author{Jonathan Halliwell}
\affiliation{Blackett Laboratory, Imperial College, London SW7 2BZ, United Kingdom}
\author{Raymond Laflamme} 
\affiliation{Institute for Quantum Computing and Department of Physics and Astronomy, University of Waterloo, Waterloo, Ontario N2L 3G1, Canada} 
\affiliation{Perimeter Institute for Theoretical Physics, Waterloo, Ontario N2L 2Y5, Canada} 

\date{\today}% It is always \today, today,

\begin{abstract}
Macroscopic realism (MR) is the view that a system may possess definite properties at any time independent of past or future measurements, and may be tested experimentally using the Leggett-Garg inequalities (LGIs). In this work we advance the study of LGIs in two ways using experiments carried out on a nuclear magnetic resonance spectrometer. Firstly, we addresses the fact that the LGIs are only necessary conditions for MR but not sufficient ones. We implement a recently-proposed test of necessary and sufficient conditions for MR which consists of a combination of the original four three-time LGIs augmented with a set of twelve two-time LGIs. We explore different regimes in which the two- and three-time LGIs may each be satisfied or violated. Secondly, we implement a recent proposal for a measurement protocol which determines the temporal correlation functions in an approximately non-invasive manner. It employs a measurement of the velocity of a dichotomic variable $Q$, continuous in time, from which a possible sign change of $Q$ may be determined in a single measurement of an ancilla coupled to the velocity. This protocol involves a significantly different set of assumptions to the traditional ideal negative measurement protocol and a comparison with the latter is carried out.
\end{abstract}
\maketitle

\section{Introduction}
Quantum technologies have shown their potential to impact a  range of sectors, including communications, finance, health and security \cite{mohseni2017commercialize}. Harnessing the power of quantum mechanics into practical quantum technologies will be aided by deeper understandings of the foundations of quantum mechanics. Without a true understanding of foundational concepts, famous quantum breakthroughs such as the BB84 protocol \cite{bennett2014quantum} and Shor's factoring algorithm \cite{shor1994algorithms} would never have been possible. A propitious question left to explore in quantum foundations is that-- If reality is described by quantum mechanics, can these laws be scaled to commonplace objects? This idea of systems composed of countless atoms existing in quantum superposition of macroscopically distinct states is known as \textit{macroscopic coherence}.

Anthony Leggett and Anupam Garg drew attention to the study of this subject by first codifying how physicist expect macroscopic objects to behave into a set of assumptions that they defined as \textit{macroscopic realism} (MR) \cite{leggett1985quantum,leggett2008realism}. As defined by Leggett and Garg these assumptions are
\begin{enumerate}
\item Macroscopic realism per se (MRps): A macroscopic system with two or more macroscopically distinct states available to it will at all times be in one or the other of these states. 
\item Noninvasive measurability (NIM):  It is possible, in principle, to determine the state of the system with arbitrarily small perturbation on its subsequent dynamics.
\end{enumerate}
Leggett and Garg later added the condition that future measurements should not affect the present state (a condition they named induction), but this assumption is rarely contested. Leggett and Garg used these assumption to derive a set of inequalities that any macroscopic system should obey. These are the Leggett-Garg inequalities (LGIs). If measurements on a system violate the LGIs then a macroscopic understanding of the system must be abandoned.  In this way the LGIs can serve as a test of macroscopic coherence. 

The violation of the LGIs on microscopic systems remains a topic of interest for different reasons. For one, violations on microscopic systems are a necessary stepping stone towards achieving macroscopic coherence. Even at the level of microscopic systems the study of the LGIs are riddled with challenges which will need to be addressed before one can feasibly move to larger systems. A second motivation for the study of the LGIs is that they serve as a test of whether a system is behaving quantum-mechanically. The use of the LGIs in this manner has been adopted in different fields including quantum transport \cite{lambert2010distinguishing}, quantum biology \cite{lambert2011macrorealism} and quantum computations \cite{morikoshi2006information}. (See Emary \textit{et al} \cite{emary2013leggett} for an extensive review of both experimental and theoretical aspects of the LG inequalities, and Ref.~\cite{maroney2014quantum} for a critique and analysis of what LGIs actually test.)

The formalism of the LGIs is fairly straightforward. First consider a dichotomic observable, $Q$, with outcomes $s_i \in \{\pm 1\}$  measured at time $t_i$. When measuring this observable $Q$ at two different times $t_i,t_j$ ($Q(t_i) \equiv Q_i)$, the outcomes will either be correlated ($s_i s_j = 1$) or anti-correlated ($s_i s_j = -1$). The classical \textit{correlation function}, $C_{ij}$, 
\begin{equation}
C_{ij} = \langle Q_i Q_j \rangle = \sum_{i,j} s_i s_j p_{ij} (s_i, s_j)
\label{eqn:defnC12}
\end{equation}
assigns a value to this correlation. $C_{ij}$ is bounded by $\pm 1$ corresponding to the cases of perfect correlation and anti-correlation respectively, and $p_{ij} (s_i, s_j)$, the \textit{two-time probability}, is the probability of obtaining the results $s_i$ and $s_j$ when measurements are made at times $t_i$, $t_j$, respectively.  By performing three experiments that measure the observable $Q$ at pairs of times $(t_1,t_2)$, $(t_2,t_3)$ and $(t_1,t_3)$ the correlation functions $C_{12}$, $C_{23}$, and $C_{13}$ can be obtained. For a system which obeys the assumptions of MRps and NIM it can be shown that these correlations are bounded by the four three-time LGIs (LG3s) \cite{leggett1985quantum, leggett2008realism}.
\begin{align}
1 + C_{12} + C_{23} + C_{13} &\geq 0 \label{eqn:LG3_1}\\
1 - C_{12} - C_{23} + C_{13} &\geq 0 \label{eqn:LG3_2}\\
1 + C_{12} - C_{23} - C_{13} &\geq 0 \label{eqn:LG3_3}\\
1 - C_{12} + C_{23} - C_{13} &\geq 0 \label{eqn:LG3_4}
\end{align}
\begin{figure}
    \includegraphics[width=\linewidth]{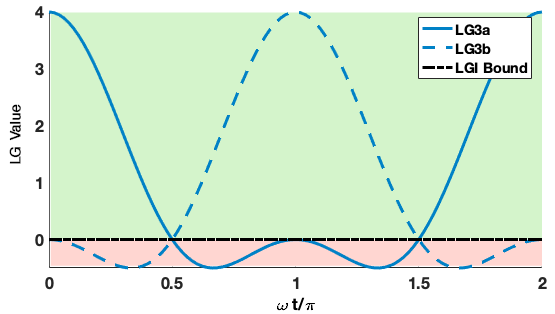}
    \caption{Two of the LG3s as functions of $\omega t /\pi$ for measurements made at equidistant time intervals. The red and green regions indicate where the LG3s are and are not, respectively, violated.}
    \label{fig:k3}
\end{figure}
These inequalities can, however, be violated by quantum systems. For example, consider a spin-$\frac{1}{2}$ particle evolving under a Hamiltonian $H =\omega \hat{X}/2$ and measured by $\hat{Q} = \hat{Z}$,  where $\hat{X}$ and $\hat{Z}$ are the Pauli-x and Pauli-z matrices. For such a model it is readily shown that for any initial state the quantum mechanical correlation function is \cite{emary2013leggett}:
\begin{align}
C_{ij} &= \frac{1}{2} \langle \hat{Q}_1 \hat{Q}_2 - \hat{Q}_2 \hat{Q}_1 \rangle \\
&= \cos(\omega(t_j - t_i)) \label{eqn:Cij}
\end{align}
By using Eq.(\ref{eqn:Cij}) and considering the case of equidistant time intervals, \textit{i.e.,} $t_j-t_i = t$, the four LG3s reduce to three inequalities
\begin{align}
(\text{LG3a})  \kern 1.0em & 1 + 2\cos(\omega t)+ \cos(2\omega t) \geq 0 \label{eqn:LG3_1s}\\
(\text{LG3b})  \kern 1.0em & 1 - 2\cos(\omega t)  + \cos(2\omega t) \geq 0 \label{eqn:LG3_2s}\\
&  \kern 5.0em 1  - \cos(2\omega t)  \geq 0 \label{eqn:LG3_3s}
\end{align}
the third of which is always satisfied. As shown in Fig.~\ref{fig:k3}, one of the other two inequalities, Eq.(\ref{eqn:LG3_1s}) or (\ref{eqn:LG3_2s}), will be violated for all but discrete choices of $\omega t$. Thus, the LG3s can be violated by a quantum system. 

In this work our aim is to advance the study of the LGIs by addressing two contemporary challenges in the field. The first concerns the question of conditions for MR that are both necessary and sufficient and the second concerns the need for LGI experiments to adopt a macroscopically non-invasive measurement protocol.

\begin{table}
\def\arraystretch{1.125}
{\setlength{\tabcolsep}{0.5em}
\begin{tabular}{ Sc | Sc | Sc | Sc |  Sc }
\textbf{Experiments} & \textbf{LG3s} & \textbf{LG2s} & \textbf{INM} & \textbf{CTVM} \\
\hline
Previously tested & \includegraphics[scale=0.05]{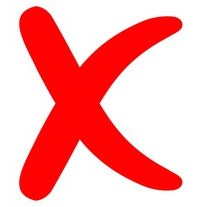} & & \includegraphics[scale=0.05]{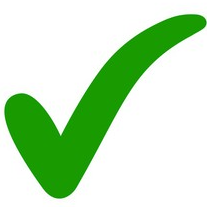} & \\
\hline 
Set 1 & \includegraphics[scale=0.05]{check} & \includegraphics[scale=0.05]{cross} & \includegraphics[scale=0.05]{check} & \\
Set 2a & \includegraphics[scale=0.05]{cross} & \includegraphics[scale=0.05]{check} & \includegraphics[scale=0.05]{check} & \\
Set 2b & \includegraphics[scale=0.05]{cross} & \includegraphics[scale=0.05]{check} &  & \includegraphics[scale=0.05]{check} \\
\end{tabular}}
\caption{An overview of the goal of each set of experiments in this work. The check marks indicate that the specified set of inequalities are all satisfied while the crosses indicate that they are violated. The check marks are also used to designate which measurement protocol is being implemented (either INM or CTVM).}
\label{tbl:Overview}
\end{table}

The first concern we address, of conditions for MR that are both necessary and sufficient, was originally addressed in Refs.~\cite{kofler2013condition, clemente2015necessary, clemente2016no} and subsequently in Refs.~\cite{halliwell2016leggett2, halliwell2017comparing}. We follow the latter papers, which concern a set of augmented LGIs in which the original LG3s are amended with a set of twelve two-time LGIs (LG2s) and form a set of necessary and sufficient conditions for MR.  Our first set of experiments demonstrates a violation of the LG2s, and hence a violation of MR not detected by the original LG framework, \textit{i.e.} the LG3s. Note that LG2s have  been considered previously in a number of experiments, but as simplifications of the LG3s, in which, for example, one sets $\langle Q_1 \rangle = 1$ as an initial condition. Here the LG2s enter in a more fundamental role, as the extra conditions required to define a set of conditions for MR which are not only necessary, but also sufficient, and are therefore a decisive test. Failure to violate the LG3s alone is not sufficient to ensure that a system can be described macrorealistically so MR tests based purely on LG3s are not fully decisive. Furthermore, the implementation of the decisive test for MR described here is clearly a desirable goal in the design of any future LG experiments.

The second challenge we look to address arises from the need of LGI experiments to adopt a macroscopically non-invasive measurement protocol. If the measurement was deemed to be invasive it could then be argued then that it was the effect of the measurement and not a failure of MR which caused the violation of the inequality \cite{montina2012dynamics, yearsley2013leggett, guhne2010compatibility}. The best one can hope to achieve in addressing this argument is to implement a measurement protocol whose argument for invasiveness would need to be so contrived that the alternative explanation of a violation of MR would be more likely. One strategy to treat this argument is to implement different measurement protocols that are constructed from different sets of assumptions. The agreement of the results from these different protocols will further strengthen either protocols argument for being non-invasive. To advance this strategy we perform in this work the first experimental implementation of the continuous in time velocity measurement (CTVM) protocol \cite{halliwell2016leggett}. The CTVM protocol is methodologically different from the more commonly implemented \cite{knee2012violation, robens2015ideal, katiyar2017experimental} technique of ideal negative measurements (INM). We implement both the CTVM and INM protocols in our second set of experiments and verify that they do provide similar results under the parameters in which the CTVM can be faithfully implemented.

Our experiments will be carried out on a nuclear magnetic resonance (NMR) spectrometer. NMR tests of the LGIs have been criticised \cite{emary2013leggett, menicucci2002local}, on the grounds that the results can always be replicated using hidden variable models. However, the models accomplishing this \cite{menicucci2002local} are of the Bohmian type (in which the state itself is one of the hidden variables) and, as stressed in Ref.~\cite{maroney2014quantum}, LG tests can rule out only certain types of hidden variable theories, and in particular, Bohmian theories can rarely be ruled out.

The content of this paper is organized into five Sections. Section \ref{sec:Theory} contains a review of the necessary theory for this work. This consists of an introduction to the augmented set of Leggett-Garg inequalities \cite{halliwell2016leggett2} and to the CTVM protocol \cite{halliwell2016leggett}. This section consists of a review of earlier works. From there, the details of the experiments performed in this work are outlined in Section \ref{sec:Experimental design}. In this section we outline why certain initial parameters were chosen, provide an overview of the different combination of experiments which are required and outline the pulse sequences for all the experiments that were implemented. The data from these experiments is then presented in Section \ref{sec:Experimental results} along with the theoretically derived and computer simulated results. This paper then concludes in Section \ref{sec:Conclusion} with a calculation of the LG2s and LG3s from the experimental data, a discussion on the different violations that had occurred and the significance of implementing the CTVM protocol.

\section{Theory} \label{sec:Theory}
\subsection{The Augmented set of Leggett-Garg Inequalities}

Despite years of experimental tests of the LGIs, the question of conditions for MR that are both necessary and sufficient has been addressed only recently \cite{kofler2013condition, clemente2015necessary, halliwell2016leggett2, halliwell2017comparing}.  The LG framework was designed in close parallel to tests of local realism using the Bell and CHSH inequalities \cite{bell1964einstein}. There, Fine's theorem \cite{fine1982hidden} ensures that the Bell and CHSH inequalities are both necessary and sufficient conditions for local realism. The LG framework differs at this point since Fine's theorem does not immediately apply and as a consequence the usual three-time LGIs are only necessary conditions for MR and not sufficient ones. The difference arises from the fact that, for pairs of measurements acting sequentially in time, the so-called ``no- signalling in time" (NSIT) conditions
\begin{equation}
p_j(s_j) = \sum_{s_i} p_{ij}(s_i, s_j) \label{eqn:NSIT}
\end{equation}
do not hold in general. Here $p_{ij} (s_i,s_j)$ is the two-time probability defined earlier and $p_j(s_j)$ is the single time probability for obtaining the result $s_j$ at time $t_j$ in which no earlier measurement is made. By contrast in Bell tests the analogous conditions are ensured by locality. As a consequence, pairwise probabilities of the form $p_{12} (s_1,s_2)$, for example, are not in general compatible with the probabilities $p_{23} (s_2,s_3)$ on their overlap. This means that Fine's theorem, which seeks an underlying joint probability matching a {\it compatible} set of marginals, does not immediately apply.

In the current literature there are two different approaches to this shortcoming. One involves a set of NSIT conditions of the form of Eq.(\ref{eqn:NSIT}) (and generalizations to three times) which simply restricts the parameter space to situations in which such conditions are satisfied \cite{kofler2013condition, clemente2015necessary}. These are quite strong conditions which, in quantum mechanics, require zero interference. The other approach, which remains close to the original LG framework, adopts an indirect procedure for determining the two-time probabilities in which the averages $\langle Q_i \rangle $, $\langle Q_j \rangle $ and the correlation function $C_{ij}$ are determined, non-invasively, in three separate experiments \cite{halliwell2016leggett2, halliwell2017comparing}. They may then be assembled into a two-time probability if and only if the following two-time LGIs hold:
\begin{equation}
1 + s_i \langle Q_i \rangle + s_j \langle Q_j \rangle + s_i s_j C_{ij} \geq 0 \label{eqn:LG2gen}
\end{equation}
where $ij$ takes the values $12$, $23$, $13$. These twelve conditions are clearly much weaker than the NSIT conditions and in quantum mechanics require only suitable bounds on the degree of interference. The two-time probabilities themselves $p(s_i, s_j)$ are then given by the left-hand side of this expression, multiplied by $\frac{1}{4}$. The indirect measurement procedure ensures that different two-time probabilities determined in this way are then compatible with each other (and so satisfy NSIT in a formal sense, but this does not say anything about signaling) and Fine's theorem then applies. We thus obtain a set of necessary and sufficient conditions for MR consisting of the four original LG3s augmented with the twelve LG2s, Eq.(\ref{eqn:LG2gen}). 

In this work, we experimentally test the definition of MR using these augmented LGIs. There are four regimes of interest, depending on whether each of the LG2s and LG3s are, or are not, satisfied. It would clearly be of interest to explore all four regimes but here we restrict to the two most interesting case. The first is the case in which the LG3s are satisfied but the LG2s are violated. This is the new regime compared to the original LG framework and detects MR violations not detected by the LG3s alone. The second is the case in which the LG2s are satisfied but the LG3s are violated. This is a natural parallel with the Bell case, in which the situation ``looks classical" for the partial snapshots consisting of the pairwise measurements, but the violation of MR is only apparent when one looks for a three-time unifying probability.

\subsection{Continuous in time velocity measurements} \label{sec:CTVMTHEORY}

A key requirement in all LGI experiments is the use of a macroscopically non-invasive measurement technique, so as to avoid the ``clumsiness loophole".  The loophole argues that a violation of the LGIs could have been caused by the measurement influencing an unknown hidden variable \cite{wilde2012addressing}. This loophole can never be entirely closed since other hidden variables can always exist \cite{emary2013leggett}. Instead, what one hopes to achieve is to implement a measurement protocol whose argument for invasiveness would need to be so contrived that the alternative explanation of a violation of MR would be more likely.

The difficulty in experimentally implementing non-invasive protocols along with the critiques of their non-invasiveness suggests that it remains of interest to implement alternative approaches. In striving for this aim we perform the first implementation of the \textit{continuous in time velocity measurement} (CTVM) protocol \cite{halliwell2016leggett} and, for comparison, also implement the current benchmark technique of ideal negative measurements (INM) \cite{leggett1985quantum}. The CTVM and INM methods are  formulated on very different sets of assumptions and so they provide different perspectives on non-invasiveness. Since these models assumptions differ so widely, their agreement on the outcome of the measured results will provide a much stronger argument for the non-invasiveness of either protocol.

Common to most methods for measuring the correlation function is the need to conduct a pair of measurements at successive times. Such models carry the potential source of invasiveness from the earlier measurement affecting the later one. The CTVM protocol avoids this feature.  It arose from the general observation that the correlation function depends only on whether $Q$ takes the same sign or opposite signs at the initial and final times \cite{halliwell2016decoherent}. This in turn depends on how many sign changes $Q(t)$ makes during the given time interval. Of course it could change sign many times in general. However, we make the simplifying assumption that in the vast majority of histories, $Q(t)$ changes sign only once. This assumption may seem like a rather restrictive one, but it has been argued that there is in fact a regime in which this assumption is reasonable \cite{halliwell2016leggett}. A single sign change can then be registered using a weakly coupled ``waiting detector”, which is designed to click if $Q(t)$ changes sign, but otherwise remains unchanged. Because this protocol involves just a single interaction at some (unknown) time during the given time interval, it is essentially non-invasive, since there are no later measurements to disturb. The only possible source of invasiveness is that the single interaction with the detector when $Q$ changes sign may cause $Q$ to change sign a second time and hence to interact with the detector a second time yielding a false detector result. However, as argued in Ref.~\cite{halliwell2016leggett}, for a weakly interacting detector, the probability for this happening is considerably smaller than the single click probability we seek.

The waiting detector is readily modelled by assuming that the primary system may be assigned a velocity $ v= \dot Q $ and then weakly coupling this to an ancilla with which it interacts continuously in time. It is readily shown that the ancilla then responds to the quantity
\begin{equation}
\int_{t_i}^{t_j}  v(t) dt = Q_j - Q_i.
\end{equation}
From this, the correlation function is then readily found from the formula,
\begin{equation}
\langle [Q_j - Q_i]^2 \rangle = 2 ( 1 - C_{ij})
\end{equation}
The existence of a velocity is an assumption stronger than what is normally supposed in LG tests (which typically take a ``black box" approach to the system and its dynamics as much as possible) but in practice LG tests are carried out on specific systems for which a velocity is readily identified. We will discuss the above two assumptions in more detail in what follows.
 
The quantum-mechanical implementation of such a protocol will require a Hamiltonian which reflects the characteristics outlined above. We consider the previously defined spin model with $H= \omega \hat{X}/2$ and operator $Q=\hat{Z}$. We also define a velocity operator $\dot Q = \omega \hat{Y}$. The total system-detector Hamiltonian then for the system (S) and ancilla (A) is
\begin{equation}
H_{D} = \frac{\omega}{2}X_{S} \otimes I_{A} + \lambda \omega Y_{S} \otimes X_{A} \label{eqn:Ham}
\end{equation}
The first term represents the evolution under the desired Hamiltonian of the system, and the second term represents the coupling of the velocity operator with $X$ on the ancilla, where $\lambda$ corresponds to the strength of the coupling. The  $X$ gate acts as a flipping operator on the ancilla when the ancilla is in the $Z$ basis. The ancilla will be initialized to the $+1$ eigenstate of $Z$ ($\ket{0}$) that flips to the $-1$ eigenstate of $Z$ ($\ket{1}$) when $Q$ changes sign. 

The value of $C_{ij}$ can be extracted from the final value of the ancilla. First note that $H_{D}^2 = (\Omega^2/4) I$, where $\Omega = \omega \sqrt{1+4\lambda^2}$. %If our system begins in the $+1$ eigenstate of $X$, $\ket{+}$, then the final state of the system after time $t$ is
From this it is easy to show that
\begin{equation}
e^{-iH_Dt} = \cos\Big(\frac{\Omega t}{2}\Big)I - \frac{2i}{\Omega} \sin\Big(\frac{\Omega t}{2}\Big)H_D
\end{equation}
The total state of the system at time $t$ is then
\begin{align*}
\ket{\Psi_t} &= e^{-iH_Dt}(\ket{\psi} \otimes \ket{0})\\
&=  \hat{A}_0(t) \ket{+}\otimes\ket{0} + \hat{A}_1(t) \ket{+}\otimes\ket{1} \label{eqn:Use4BA}
\end{align*}
for 
\begin{align}
\hat{A}_0(t) &= \cos\Big(\frac{\Omega t}{2}\Big)I_S - \frac{i\omega}{\Omega}\sin\Big(\frac{\Omega t}{2}\Big)X_S\\
\hat{A}_1(t) &= - \frac{2i\lambda\omega}{\Omega} \sin\Big(\frac{\Omega t}{2}\Big)Y_S
\end{align}
thus the probability of the ancilla being in the state $\ket{1}$ after a time $t$ evolution is
\begin{align}
p(1) &= \bra{+} \hat{A}_1(t)^{\dagger}\hat{A}_1(t) \ket{+} \label{eqn:p(1)}\\
&= \frac{2\lambda^2\omega^2}{\Omega^2} (1 - \cos(\Omega t))
\end{align}
For a sufficiently small $\lambda$
\begin{align}
p(1) &\approx 2\lambda^2 (1-\cos(\omega t))\\
&= 2\lambda^2(1 - C_{12}) \label{eqn:cijp(1)}
\end{align}
Thus, $C_{ij}$ can be calculated with a single measurement that determines the probability of $Q$ changing signs over the time interval $[t_i, t_j]$. 

\subsection{Two Regimes of Interest} \label{sec:2regimes}

In this work we carry out two sets of experiments which explore the two regimes of LGI violations of greatest interest (see Table \ref{tbl:Overview}). In both sets of experiments the requirement of non-invasiveness is primarily accomplished by using ideal negative measurements. However, in the second set of experiments the CTVM protocol is also implemented for comparison.

\textit{First set of experiments --} As mentioned, the goal of the first set of experiments is to demonstrate a violation of the LG2s while the LG3s are satisfied. We will first consider the requirement of satisfying the LG3s. As was shown in Fig.~\ref{fig:k3}, for the case of equidistant time intervals, the LG3s will only be satisfied for $\omega t = n \frac{\pi}{2}$ for $n \in \mathbb{Z}$. Still for the case of non-equidistant time intervals, the LG3s are still only satisfied for discrete values of $\omega t$. Thus, to experimentally satisfy the LG3s we will require a means of widening the range of values of $\omega t$ in which the LG3s are satisfied.

\begin{figure}
    \includegraphics[width=\linewidth]{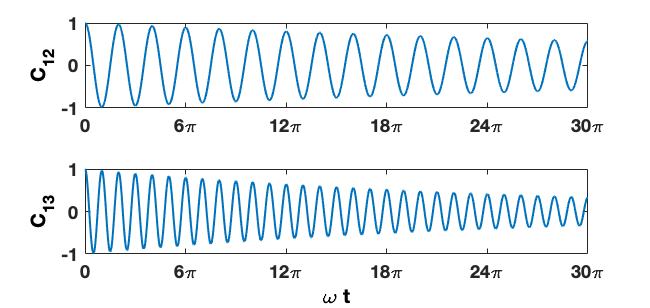}
    \caption{A simulation of the effect of the decoherence on the correlation functions as a function of $\omega t$.}
    \label{fig:decoCij}
\end{figure}

\begin{figure}
    \includegraphics[width=\linewidth]{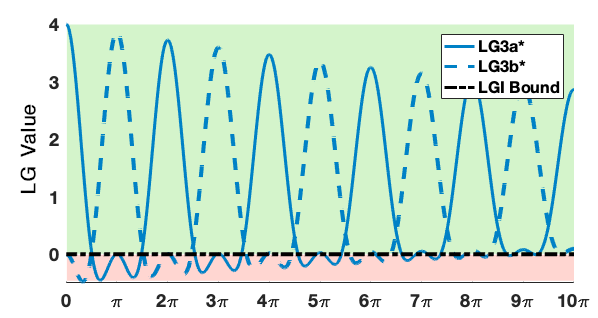}
    \caption{Eq.(\ref{eqn:LG3_1s}) and (\ref{eqn:LG3_2s}) are plotted with the dampening effect of the decoherence. As $\omega t$ increase the LG3s gradually have larger regions in which they exist above the LGI bound. The decoherence in this figure is exaggerated for clarity.}
    \label{fig:decoViol}
\end{figure}

As shown by Athalye et al \cite{athalye2011investigation}, this may be accomplished by taking advantage of the small amount of decoherence naturally present in the system due to the unavoidable interactions with the surroundings. The decoherence dampens the magnitude of the correlation functions with time as shown in Fig.~\ref{fig:decoCij}. This will in turn dampen the LG3s (Eq.(\ref{eqn:LG3_1s}) and (\ref{eqn:LG3_2s})) as shown in Fig.~\ref{fig:decoViol}. This gradual dampening leads to progressively larger ranges of $\omega t$,  centred around multiples of $\omega t = \frac{\pi}{2}$,  in which the LG3s are satisfied (above the LGI bound). An alternative approach here would be to explore definitions of MR which work with the four-time LG inequalities \cite{halliwell2017comparing} (LG4s), instead of the LG3s, since these do in fact have non-trivial regimes in which all LG4s are satisfied (although involve more measurements). This will be explored elsewhere. The next question is then -- Does there exist parameters in these ranges of $\omega t$ in which the LG2s can still be violated even when accounting for the effect of the decoherence on the LG2s?

\begin{figure}
    \includegraphics[width=\linewidth]{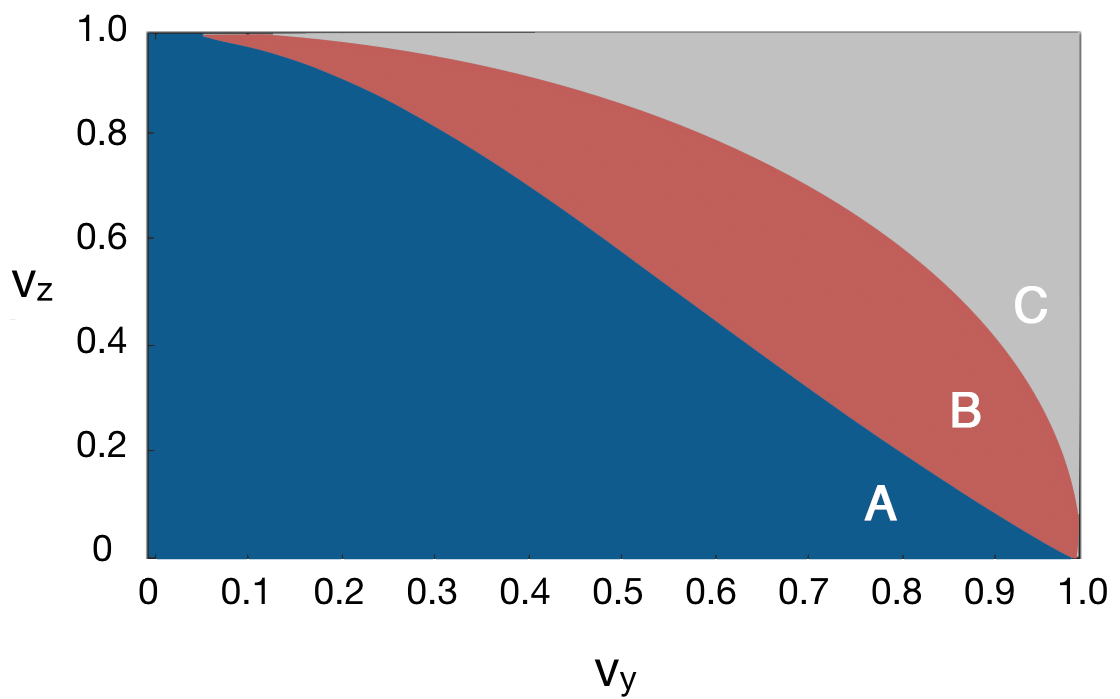}
    \caption{The results of a search over the values of $v_y, v_z$ for which all LG2s are, or are not, satisfied for $\omega t = \pi/2$. The blue region (dark grey labelled A) depicts the initial states in which the LG2s are all satisfied, the red region (grey labelled B) the states in which the LG2s are not satisfied and the grey (light grey labelled C) depicts initial states which do not exist (\textit{i.e.} outside the bound of $v_y^2 + v_z^2 \leq 1$).}
    \label{fig:searchrho1}
\end{figure}

Unlike the LG3s, the LG2s also depend on the initial state, $\rho = \frac{1}{2}(I + \vec{v}\cdot\vec{\sigma})$, in addition to the value of $\omega t$. This is due to the LG2s (Eq.(\ref{eqn:LG2gen})) being functions of $\langle  Q_i \rangle$. For our spin model the $\langle Q_i \rangle$ is equal to
\begin{align}
\langle  Q_i \rangle &= \text{tr} \big[e^{iHt_i}Ze^{-iHt_i}\rho\big] \\
& = \text{tr}\big[ Ze^{-i\omega t_iX}\rho\big]\\
&= \text{tr}\big[(\cos(\omega t_i)Z + \sin(\omega t_i)Y)\frac{1}{2}(I + \vec{v}\cdot\vec{\sigma})\big]\\
&= v_z\cos(\omega t_i) + v_y \sin(\omega t_i) \label{eqn:Qigen}
\end{align}
Thus, for any choice of $\omega t$  one can search over all possible initial states to find parameters in which the LG2s are satisfied. The results of such a search for $\omega t = \pi/2$ are presented in Fig.~\ref{fig:searchrho1}.  Of the possible initial states generated from such a search, we choose one which is experimentally simple to prepare. The state
\begin{equation}
\rho_1  = \frac{1}{2} \Big(I + \frac{Y}{\sqrt{2}} +  \frac{Z}{\sqrt{2}} \Big) \label{eqn:rho1}
\end{equation}
satisfies these criteria.

\begin{figure}
    \includegraphics[width=\linewidth]{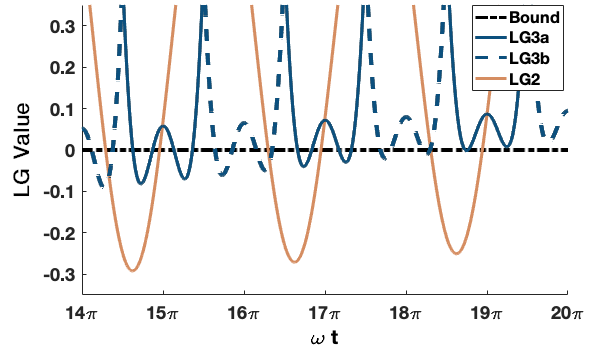}
    \caption{One of the twelve LG2s and the LG3s  (Eq.(\ref{eqn:LG3_1s}) and (\ref{eqn:LG3_2s})) are plotted as functions of $\omega t$ with the effect of the decoherence. The figure shows that there exists a region of $\omega t$ in which at least one of the LG2s is violated (exists below the bound) while the LG3s are satisfied (exist above the bound). This figure uses the same decoherence parameter as Fig.~\ref{fig:decoViol}.}
    \label{fig:dampViolLg2}
\end{figure}

Having chosen an initial state $\rho_1$ and $\omega t = \pi/2$, the effect of the decoherence on the LG2s and the LG3s can be simulated\cite{laforest2008error}. A segment of the results of such a simulation are provided in Fig.~\ref{fig:dampViolLg2}. Fig.~\ref{fig:dampViolLg2} depicts the existence of a regime in which the LG2s are violated (exist below the LGI bound) and the LG3s are satisfied (exist above the bound). In this work we use a delay of $0.1$s between each measurement interval to achieve the desired dampening effect. The details of how this is achieved is provided in Section \ref{sec:Experimental design}.

\textit{Second set of experiments --} The second set of experiments have two goals. The first is to demonstrate a violation of the LG3s while the LG2s are satisfied. As discussed, the LG3s are always violated except for discrete choices of $\omega t$. Of the possible values of $\omega t$ available to us we will choose, as will be justified in Section \ref{Experiment_CTVM}, $\omega t = 3\pi/10$. As was done for the previous set of experiments, a search is performed over the possible values of $v_y, v_z$ to find the regions in which the LG2s are all satisfied. Again, for experimental ease we will select one that is experimentally simple to prepare
\begin{equation}
\rho_2 =  \frac{1}{2}\Big(I + (0.951)X + (0.309)Z\Big) \label{eqn:rho2}
\end{equation}
This choice of initial state and $\omega t$ will provide a set of initial conditions which will violate the LG3s and satisfy the LG2s as desired.

The second sets of experiments has the additional goal of implementing the CTVM protocol. This was not done in the first set of experiments since the CTVM protocol requires a much smaller value of $\omega t$ than those available to choose from in the first set of experiments. In the second set of experiments the $\omega t$, as we show in Section \ref{Experiment_CTVM},  is sufficiently small for implementing the CTVM protocol. Thus in the second set of experiments the CTVM and INM protocols are both implemented and a comparison of the results  is performed.

\section{Experimental design} \label{sec:Experimental design}

All the experiments performed in this work are carried out at 298K on a Bruker DRX spectrometer with a nominal $^{1}$H frequency of 700 MHZ. The NMR sample consisted of $^{13}$C-chloforom dissolved in acetone to produce a heteronuclear two-spin system. The $^{1}$H was used as the ancilla qubit and the $^{13}$ C was used as the primary qubit. Both spins were placed on resonance so that the Hamiltonian consisted of only the spin-spin coupling which has a value of $215.15$ Hz. The measured relaxation times were $T_1 = 6.63$s and $T_2 = 0.76$s for $^{1}H$ and $T_1 = 8.66$s and $T_2 = 1.10$s for $^{13}C$. An inter-scan delay of $90$s was used to ensure that the spins began each experiment close to their thermal state.

Furthermore, a point can be made more clear regarding the primary qubit in question. In NMR experiments it is not guaranteed, nor is it necessary, that the ``same" collection of nuclei are to be measured in each run of the experiment. In NMR experiments the nuclei are distinguished by their Larmor frequencies. Since, all the $^{13}C$ nuclei in this sample have identical Larmor frequencies they are thus indistinguishable from one another. As such, the fair sampling assumption in these terms is justified.

\subsection{Ideal negative measurement} \label{sec:INMTheory}

In both the first and second set of experiments we require implementing the INM protocol. In the INM protocol the ancilla is coupled to only one of the two measurement outcomes.  If an experiment is performed and the ancilla changes states, then those measurements are discarded. If the ancilla does not change states then it can be inferred that the system was in the orthogonal space and those results are kept. This entire procedure is then repeated with the ancilla being coupled to the other measurement outcome. This protocol thus provides a macroscopic argument for the system-ancilla interaction not being a potential source of invasiveness. 

Before describing the full details of the INM, we first consider a general qubit that is evolving in time and is measured along $\hat{Z}$ at times $t_i, t_j$.  This system begins in some initial state $\rho$ at time $0$ and evolves freely for time $t_i$. The state $\rho$ at time $t_i$  can be written generally as
\begin{equation}
\rho_i = \begin{bmatrix} a & b \\ b^* & 1 \minus a \end{bmatrix}
\end{equation}
\begin{figure} 
    \includegraphics[width=\linewidth]{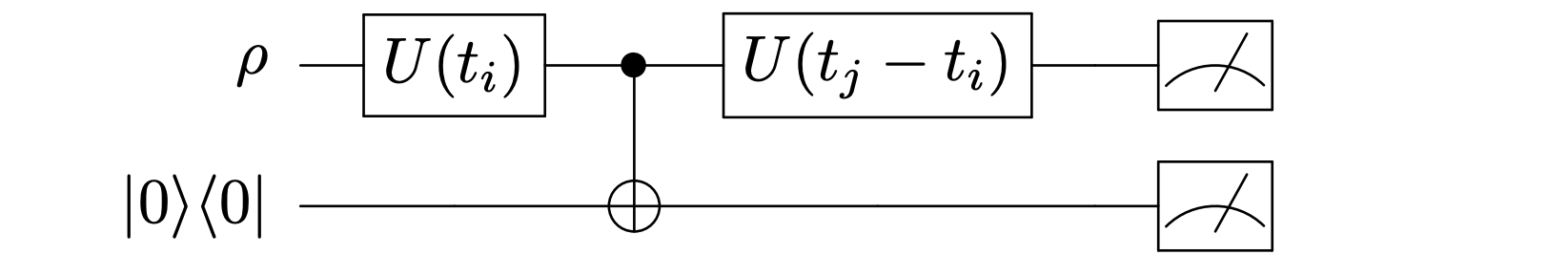}
    \caption{The quantum circuit used for implementing the INM protocol. The evolutions $U(t_i) = e^{-iHt_i}$ and  $U(t_j - t_i) = e^{-iH(t_j - t_i)}$ behave exactly as defined in Section \ref{sec:INM}.}
    \label{fig:INMcirc}
\end{figure}
where $a$ is real, $b$ is complex and $ a(1-a) \geq  |b|^2 $ (with the equality for a pure state). At this time the first measurement of $Z$ is conducted and the states $\ket{0}$ and $\ket{1}$ will be returned with probabilities $a$ and $(1 \minus a)$ respectively. Once the measurement is completed the state will update to $\ket{0}\bra{0}$ or $\ket{1}\bra{1}$ depending on the measured out come. The system then evolves freely again until time $t_j$. This second evolution can be written generally as the mapping between states
$$\ket{0}\bra{0} \rightarrow \begin{bmatrix} a' & b' \\ b'^* & 1\minus a' \end{bmatrix}, \quad \ket{1}\bra{1} \rightarrow  \begin{bmatrix} a'' & b'' \\ b''^* & 1\minus a'' \end{bmatrix}$$

At this point the second measurement of $Z$ is conducted. If the measurement outcome at $t_i$ was $\ket{0}$ then the states $\ket{0}$ and $\ket{1}$ will be returned at $t_j$ with probabilities $a'$ and $(1 \minus a')$ respectively. If the measurement outcome at $t_i$ was $\ket{1}$ then the states $\ket{0}$ and $\ket{1}$ will be returned at $t_j$ with probabilities $a''$ and $(1 \minus a'')$ respectively. Thus the two-times probabilities $p_{12}(s_i, s_j)$ for all possible measurements are equal to
\begin{align*}
&p_{12}(+,+) = aa', \qquad \quad p_{12}(+,-) =  a(1\minus a')\\
&p_{12}(-,+) = (1 \minus a)a'', \quad p_{12}(-,-) =  (1\minus a)(1 \minus a'')
\end{align*}
and, using Eq.(\ref{eqn:defnC12}),  the correlation function can be written as
\begin{align}
C_{ij} = aa' - a(1\minus a') -  (1 \minus a)a'' + (1\minus a)(1 \minus a'') \label{eqn:Cijprob}
\end{align}
It is shown in Appendix \ref{App:MeasuringCij} that the diagonal entries of the final output of the circuit in Fig.~\ref{fig:INMcirc} are  precisely $aa', a(1\minus a'), (1\minus a)a''$ and $(1\minus a)(1\minus a'')$. Thus a single measurement of these diagonal entries can be done to determine $C_{ij}$. \label{sec:INM}

This procedure to measure $C_{ij}$ can be modified to implement the INM protocol.  From a macrorealistic perspective the primary system in Fig.~\ref{fig:INMcirc} would be completely undisturbed before being measured, were it not for the potential interactions with the CNOT gate. However, from this perspective the ancilla does not interact with the CNOT gate if the primary system is in the state $\ket{0}$. Therefore, to implement a non-invasive protocol the experiment is run twice. Once the experiment is run with the CNOT gate and only the results in which the primary system was in the state $\ket{0}$ are kept ($p_{12}(+,+)$ and $p_{12}(+,-)$). Then the experiment is run again with an anti-CNOT gate and only the results where the primary system was in the state $\ket{1}$ are kept ($p_{12}(-,+)$ and $p_{12}(-,-)$). Together these two experiments provide all the information necessary to determine $C_{ij}$. Appendix \ref{App:TheoResultINM} shows that the numerical result from this measurement procedure matches the theoretical result of $C_{ij} = \cos(\omega (t_j - t_i))$ for the simple spin model.
\subsection{Continuous in time velocity measurements} \label{Experiment_CTVM}
As mentioned, the second set of experiment will also implement the CTVM protocol. The successful implementation of this protocol will require selecting values of $\lambda$ and $\omega t$ which 
\begin{enumerate}
    \item minimize the error from multiple sign changes of $Q$
    \item justify the approximation made to determine $C_{ij}$
    \item minimize the effect of the back action
    \item produce a detectable signal
\end{enumerate}
The first condition depends only on the choice of $\omega t$, the second condition depends only on the choice of $\lambda$ and the last two conditions will depend on both. Our objective in this Section is to present a procedure which identifies a choice of parameters that suitably minimize the sources of error to faithfully implement the CTVM protocol.

\textit{Single sign change--}
Given that the LGIs are designed to rule out certain types of hidden variables models, we need to assess the assumption of a single sign change of $Q$ from that perspective as well as from a quantum mechanical one. As shown in Ref.~\cite{halliwell2016leggett}, in a simple hidden variable model, the value of $Q(t)$ is determined by the direction of a unit vector rotating around a single axis. Our system is evolving under the Hamiltonian $H = \omega X/2$, i.e. the vector representing the state is rotating with frequency $\omega$ around the $x$-axis. If the vector lies in the half of the hemisphere corresponding to $Q = +1$ then the vector can rotate into the opposite hemisphere but not come back out if the total time of the evolution is less than $\pi/\omega$.  Since the measurements are made at regular time intervals, $t$ $(t_3 - t_2 = t_2 - t_1 = t)$, then the longest time which we require only one sign change to occur in is $2t$ (for determining $C_{13}$). Thus in a simple hidden variable description of the system there will be a maximum of one sign change if $\omega t \leq \pi /2$. 
\begin{figure}
    \includegraphics[width=\linewidth]{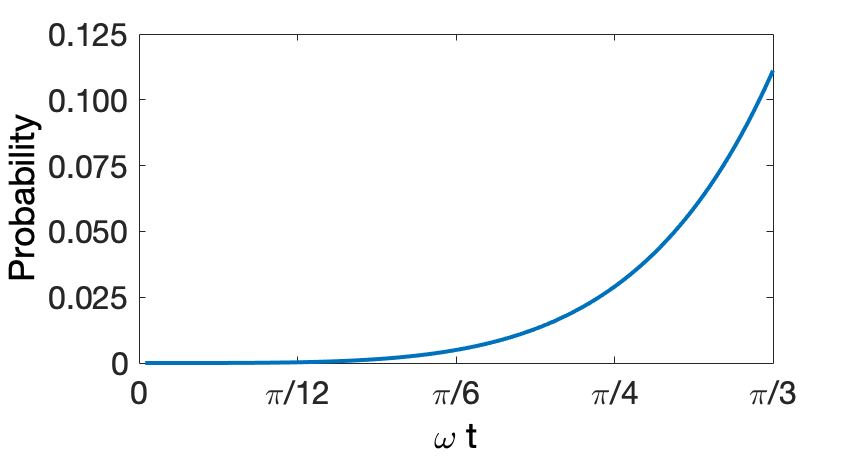}
    \caption{The probability of $Q$ undergoing multiple sign changes as a function of $\omega t$.}
    \label{fig:mergedfigs}
\end{figure}

Since we expect our system to conform to the laws of quantum mechanics, we can use a quantum model to determine the fraction of histories where $Q$ will have two sign changes. By using $H = \omega X/2$, $Q = Z$ and defining the $Z$ eigenstates by $\ket{\pm}$, the probability that $Q$ takes values of  $+1, -1, +1$ at times $0, t, 2t$ is
\begin{align}
p_{123}(+,-,+) &= | \bra{+}e^{-iHt}\ket{-}|^2  | \bra{-}e^{-iHt}\ket{+}|^2\\
&= \sin^4\Big(\frac{\omega t}{2} \Big) \label{eqn:p+-+}
\end{align}
and similarly the probability that $Q$ takes the value of $+1$ at all times to be
\begin{equation}
p_{123}(+,+,+) = \cos^4\Big(\frac{\omega t}{2} \Big) \label{eqn:p+++}\\
\end{equation}
which gives us the ratio of paths with two sign changes to paths with none as
\begin{equation}
\frac{p_{123}(+,-,+)}{p_{123}(+,+,+)} = \tan^4\Big(\frac{\omega t}{2} \Big) \label{eqn:multisign}
\end{equation}
This probability is plotted as a function of $\omega t$ in Fig.~\ref{fig:mergedfigs}.

\textit{$C_{ij}$ approximation--} In the derivation of the correlation functions an approxmation is necessarily made that 
\begin{equation}
\sqrt{1+4\lambda^2} \approx 1 \label{eqn:CijApprox}
\end{equation}
The degree of the accuracy of this approximation can be considered as another probability of error. To stay true to the spirit of the CTVM protocol we will need to choose a $\lambda$ for which this approximation is reasonable.
\begin{figure}
    \includegraphics[width=\linewidth]{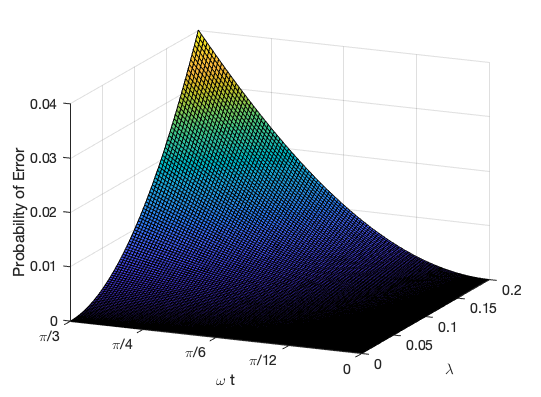}
    \caption{The probability of error from the back action of the ancilla on the system as a function of $\lambda$ and $\omega t$.}
    \label{fig:probbackaction}
\end{figure}

\textit{Back action--} As mentioned before we must consider the potential effect of the back action from the ancilla on the system. Consider the probability that an experiment is conducted for time $2t$ where the expected result for the ancilla is $\ket{1}$ (\textit{i.e.} one sign change) but due to the effect of back action the detected result is $\ket{0}$ .
The probability that the ancilla is in the state $\ket{0} \rightarrow \ket{1} \rightarrow \ket{0}$ at times $0, t, 2t$ follows from Eq.(\ref{eqn:Use4BA}) to be
\begin{align}
P_{\ket{0} \rightarrow \ket{1} \rightarrow \ket{0}} &= \bra{\psi }(\hat{A}_1(t)^{\dagger}\hat{A}_1(t))^2 \ket{\psi} \\
&= 16 \lambda ^4 \sin^4\Big(\frac{\omega t}{2} \Big)
\end{align}
similarly the probability that the ancilla is in the state $\ket{0} \rightarrow \ket{1} \rightarrow \ket{1}$ at times $0, t, 2t$ is
\begin{align}
P_{\ket{0} \rightarrow \ket{1} \rightarrow \ket{1}} &= \bra{\psi}(\hat{A}_0(t)^{\dagger}\hat{A}_1(t))^2 \ket{\psi} \\
&= 4 \lambda^2 \sin^2 \Big(\frac{\omega t}{2} \Big)
\end{align}
the ratio of these probabilities is
\begin{align}
\frac{P_{\ket{0} \rightarrow \ket{1} \rightarrow \ket{0}}}{P_{\ket{0} \rightarrow \ket{1} \rightarrow \ket{1}}} =  4 \lambda^2 \sin^2 \Big(\frac{\omega t}{2} \Big) \label{eqn:ProbBA}
\end{align}
This probability of error from the back action (Eq.(\ref{eqn:ProbBA})) is plotted as a function of $\lambda$ and $\omega t$ in Fig.~\ref{fig:probbackaction}.

\begin{figure}
    \includegraphics[width=\linewidth]{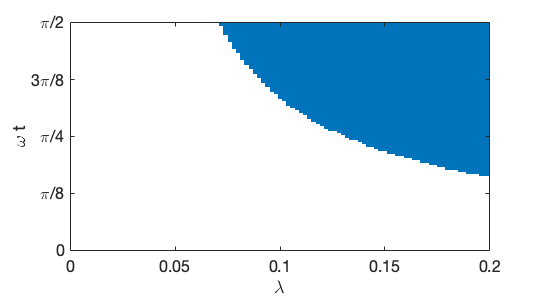}
    \caption{The values of $\lambda$ and $\omega t$ for which the value of $p(1)$ from Eq.(\ref{eqn:p1t}) is greater than $0.01$ are highlighted in blue.}
    \label{fig:allowedp1}
\end{figure}

\textit{Detectable signal--} For the last two conditions that we considered it was most favourable to minimize $\lambda$ to the furthest degree possible. Unfortunately, as $\lambda$ decreases so to will the probability of measuring the value of $p(1)$ that is required for determining $C_{ij}$. If $p(1)$ is too small then it can not be accurately measured. The error on the measurement can be limited to occur on the third decimal place with the use of multiple scans, so we will restrict $p(1) \geq 0.01$. Since experiments are conducted at both time $t$ and $2t$ we need to consider two values of $p(1)$
\begin{align}
p(1)_{t} &=\bra{1}(\text{Tr}_1(e^{-iHt}\rho e^{iHt}))\ket{1} \label{eqn:p1t}\\
p(1)_{2t} &=\bra{1}(\text{Tr}_1(e^{-2iHt}\rho e^{i2Ht}))\ket{1} \label{eqn:p2t}
\end{align}
It is easy to check algebraically that Eq.(\ref{eqn:p1t}) is strictly less than Eq.(\ref{eqn:p2t}). Intuitively this is also straightforward, since a longer time of coupling between the system and the ancilla can only lead to a greater probability of flipping the ancilla. Thus we only need to consider the lower bound on Eq.(\ref{eqn:p1t}). The values of $\lambda$ and $\omega t$ in which Eq.(\ref{eqn:p1t}) is greater than $0.01$ are plotted in Fig.~\ref{fig:allowedp1}.

Our approach in addressing these different conditions was to first prioritize justifying the approximation from Eq.(\ref{eqn:CijApprox}) to stay true to the spirit of the initial CTVM proposal.  As shown in Fig.~\ref{fig:allowedp1} as our choice of $\lambda$ decreases our choice of $\omega t$ must subsequently increase to maintain having a detectable signal. Subsequently, as depicted in Fig.~\ref{fig:mergedfigs} our value of $\omega t$ can only increase so much before the error of multiple sign changes becomes too large. Lastly, as was mentioned the probability of error from the back action remains significantly smaller than the other sources of error for the range of  $\omega t$ and $\lambda$ which are feasible. With these considerations in mind we choose a $\lambda = 0.11$ and $\omega t = 3\pi/10$. This provides a probability of error from multiple sign changes of $\approx 0.067$, a probability of error from the back action of $\approx 0.01$, the approximation of $\sqrt{1+4\lambda^2}  = 1$ being satisfied to with in $.02$ and the production of a detectable signal. 

\begin{figure}
    \includegraphics[width=\linewidth]{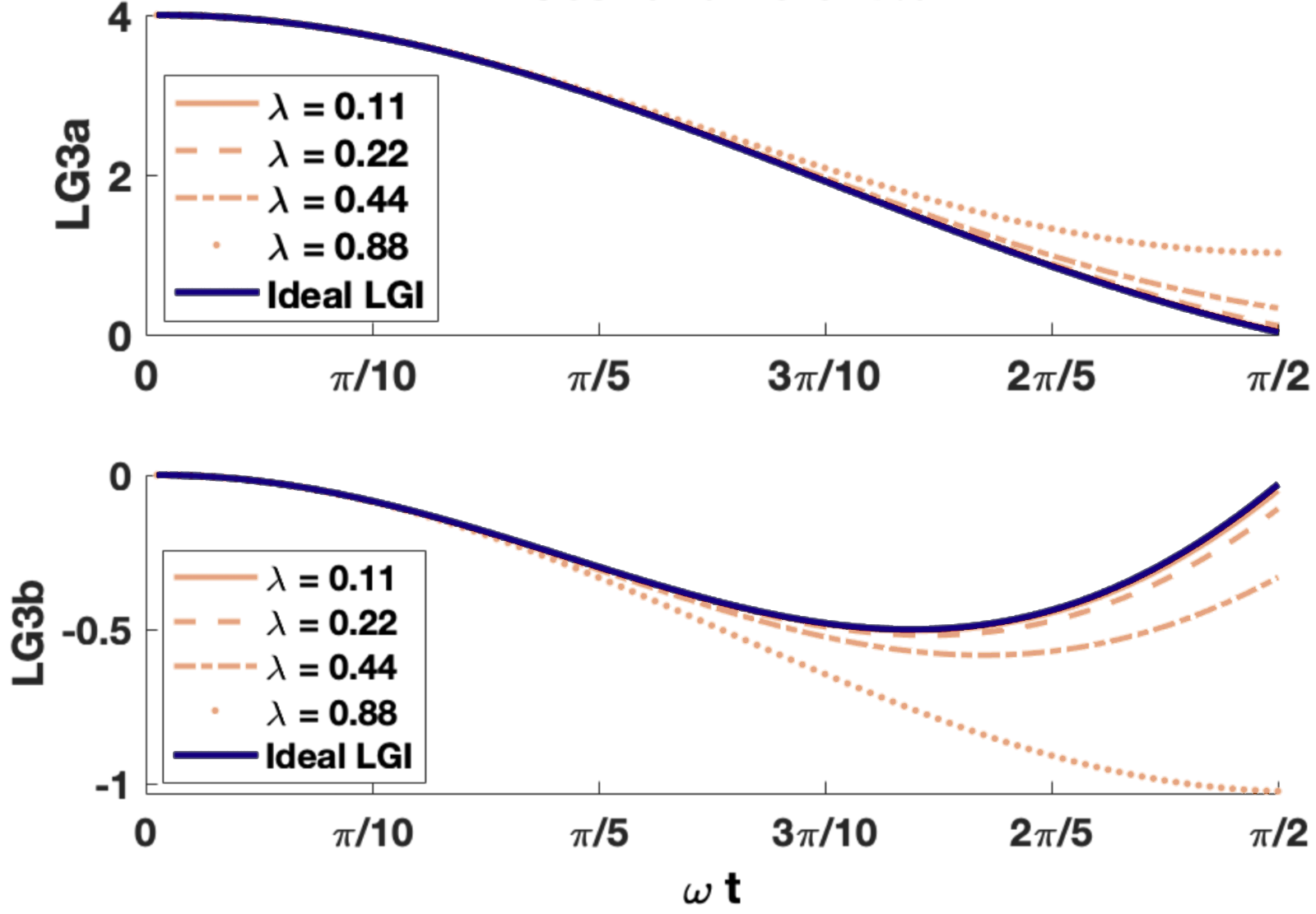}
    \caption{The effect of the different values of $\lambda$ on the LG3s.}
    \label{fig:lambdaCij}
\end{figure}

\textit{Bounds on the Violation--} A final consideration that we must take is the effect of our choice of parameters on the LG3 violation.  As noted, the derivation of the correlation functions with the CTVM protocol requires the approximation $\sqrt{1+4\lambda^2} \approx 1$. The theoretical values for $C_{ij}$ from the CTVM protocol will differ to some extent from the ideal values of $C_{ij} = \cos(\omega t)$. The larger this difference is the greater of a violation of the LG3s we must have for the source of this violation to not be caused by the approximation. Fig.~\ref{fig:lambdaCij} compares the LG3s constructed using the ideal $C_{ij}$ with those constructed using the theoretical correlation functions of the CTVM protocol for different choices of $\lambda$. For the first LG3 in Fig.~\ref{fig:lambdaCij} the greater values of $\lambda$ create larger violations of the inequality but for the second LG3 the opposite is true. Thus, we only need to worry about potential violations coming from non-zero $\lambda$ for the first case. For our choice of $\lambda = 0.11$ and $\omega t = 3\pi/10$ the LG3 is $0.0028$ less than the ideal value. So to have a violation of the LG3s we will need to use a bound of $-0.0028$ instead of $0$. 

The analysis of the possible errors coming from more than one sign change or from back reaction are purely theoretical estimates. However, given that the single sign change assumption is key to the CTVM method, it would clearly be of interest to check it using a set of control experiments in which the relevant quantities such as Eqs.\eqref{eqn:multisign} and \eqref{eqn:ProbBA} are determined experimentally. We do not do this here but it will be pursued in future works.

\begin{figure*}
\begin{center}
    \includegraphics[width=\linewidth]{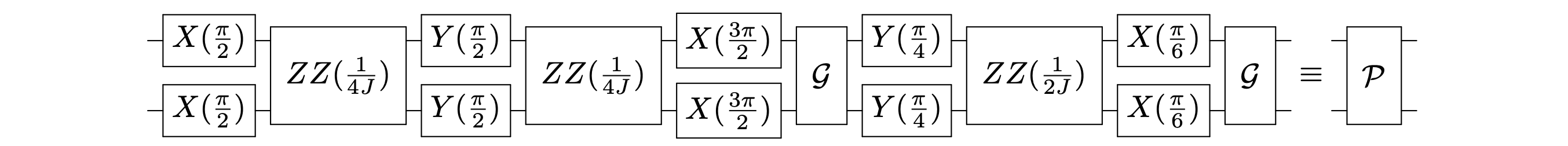}
    \caption{The pulse sequence used for producing the pseudo-pure state. $X(n)$ and $Y(n)$ depict rotations of $n$ radians around the $X$ and $Y$ axis ($X(n) = e^{-iX\frac{n}{2}}$ and $Y(n) = e^{-iY\frac{n}{2}}$). $ZZ(n)$ depicts the free evolution of the system that will provide an $n$ radian rotation of $ZZ$ ($ZZ(n) = e^{-iZZ\frac{n}{2}}$) and $\mathcal{G}$ represents the application of a gradient.}
    \label{fig:pps}
\end{center}
\end{figure*}

\subsection{Pulse sequences} \label{sec:components}
In this subsection we will outline the pulse sequences for the two sets of experiments. In total, this will consist of twenty-one different pulse sequences that are each composed of a combination of twenty to thirty different individual pulses, free evolutions and gradients. We first group together different pulse sequences into components and label these components according to their function. We then use these components to construct pulse sequences which are easier to interpret.

\begin{figure}
    \includegraphics[width=\linewidth]{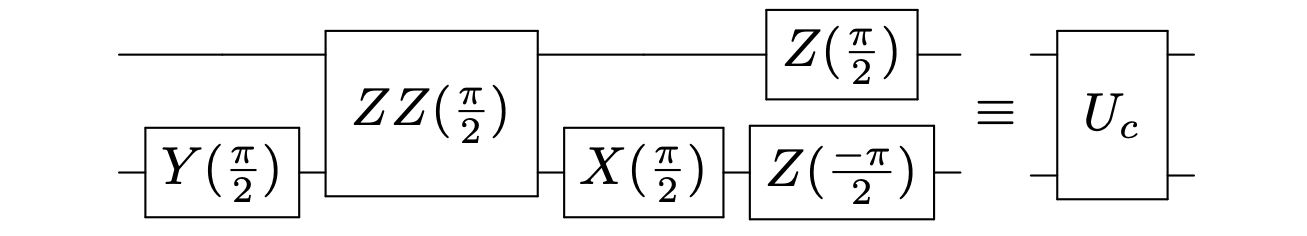}
    \caption{A pulse sequence for implementing the CNOT gate in NMR. The full sequence will be refereed to as $U_c$.}
    \label{fig:CNOT}
\end{figure}

\subsubsection{Components}
The first component of each experiment will be a pulse sequence for preparing the pseudo-pure state (pps)  \cite{cory1997ensemble, gershenfeld1998quantum}. The pps is the analog of the $\ket{0}^{\otimes n}$ state for NMR experiments. The pulse sequence used to prepare the pps in this work is given in Fig.~\ref{fig:pps} and we define this component as $\mathcal{P}$. Preparing the pps is the first step in preparing our desired initial states. Recall in Section \ref{sec:2regimes} that we chose to use pure initial states, this was done since a pure state can be prepared from the pps with a single pulse. Preparing the state $\rho_1$ (Eq.(\ref{eqn:rho1})),  for the first set of experiments, requires preparing the pps and then performing a $X(\frac{-\pi}{4})$ rotation,  we define this entire procedure as
\begin{equation}
\mathcal{P}_1 \equiv (\mathcal{P})(X(\tfrac{-\pi}{4}) \otimes I)
\end{equation}
Likewise, preparing the state $\rho_2$ (Eq.(\ref{eqn:rho2})), for the second set of experiments,  requires preparing the pps and then performing a $Y(\frac{2\pi}{5})$, we define this entire procedure as
\begin{equation}
\mathcal{P}_2 \equiv (\mathcal{P})(Y(\tfrac{2\pi}{5}) \otimes I)
\end{equation}
Thus the components $\mathcal{P}_1$ and $\mathcal{P}_2$ represent the full preparation procedures for the first and second set of experiments respectively.

\begin{figure}
    \includegraphics[width=\linewidth]{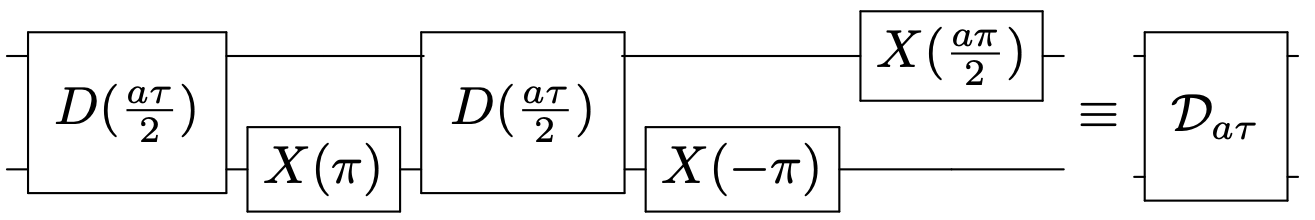}
    \caption{A pulse sequence for implementing the time delay while undoing the effect of the j-coupling.  $D(\frac{a\tau}{2})$ depict waiting for time $\frac{a\tau}{2}$. The full sequence will be refereed to as $\mathcal{D}_{a\tau}$.}
    \label{fig:waitcirc}
\end{figure}

The implementation of the INM protocol broadly consists of two components. The first is the CNOT and anti-CNOT gates. The pulse sequence used to implement the CNOT in this work is given in Fig.~\ref{fig:pps} and we define this component as $U_{c}$. The anti-CNOT can be performed by implementing $U_{c}$ with a preceding and succeeding $X(\pi)$ on the first qubit. We define this sequence for implementing the anti-CNOT as
\begin{equation}
U_{ac} \equiv  (X(\pi) \otimes I )(U_{c})(X(\pi) \otimes I )
\end{equation}
The second component for the INM protocol consists of the evolution of the system between measurements. For a system with a Hamiltonian $H = \omega X$ being measured at equidistant time intervals the system will undergo a $X(\omega t)$ or $X(2\omega t)$ evolution between measurements. These evolutions can be implemented with a single Pauli rotation and this is how we implement these evolutions in the second set of experiments. However, in the first set of experiments we also require the system to experience the effect of the decoherence as well during the evolution between measurements. This is done by implementing a time delay. During the time delay the system will also evolve according to its natural Hamiltonian. Since both spins are placed on resonance this natural Hamiltonian will consist of only the $j$-coupling term. We undo the effect of the $j$-coupling through the use of $\pi$-pulses. After the decoherence is implemented in this fashion for a time $\tau$, a Pauli rotation can then be used to implement the  $X(\omega t)$ or $X(2\omega t)$ evolution. This entire component is depicted in Fig.~(\ref{fig:waitcirc}) and is labelled as $\mathcal{D}_{a\tau}$.

\begin{figure}
    \includegraphics[width=\linewidth]{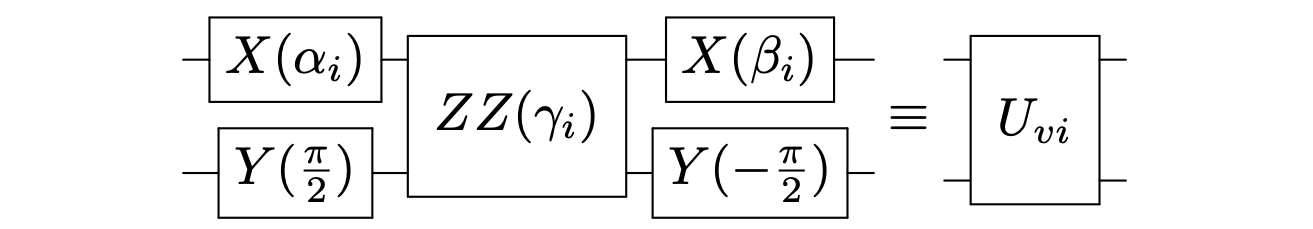}
    \caption{The pulse sequence used to implement the system-detector evolution. The parameters $\alpha_1 = 4.9751$, $\beta_1 = 1.8335$ and $\gamma_1 = 0.1035$ are used for implementing $U_{v1}$ and the parameters $\alpha_2 = 5.2433$, $\beta_2 =  2.1018$ and $\gamma_2 = 0.1998$ are used for implementing $U_{v2}$.}
    \label{fig:simhamcirc}
\end{figure}

Additionally, the pulse sequence for implementing the CTVM protocol will also require its own components. Besides the preparation $\mathcal{P}_2$ it will also require a component for implementing the system-detector set up for times $t$ and $2t$. The Hamiltonian for the system-detector evolution was given in Eq.(\ref{eqn:Ham}) as $H_D$. Thus, implementing the system-detector evolution requires constructing a pulse sequence whose full evolution is equal to $U_{v1} = e^{-iH_Dt}$ for the coupling of time $t$ and another whose evolution is equal to $U_{v2} = e^{-2iH_Dt}$ for the coupling of time $2t$. The pulse sequences for implementing $U_{v1}$ and $U_{v2}$ are provided in Fig.~\ref{fig:simhamcirc}. It is readily shown that the full evolution of these systems is equal to $e^{-iH_Dt}$ and $e^{-2iH_Dt}$ respectively. 

The NMR pulse sequences for the first and second set of experiments can now be constructed using only the components defined in this section.

\subsubsection{First set of experiments}
\begin{figure}
    \includegraphics[width=\linewidth]{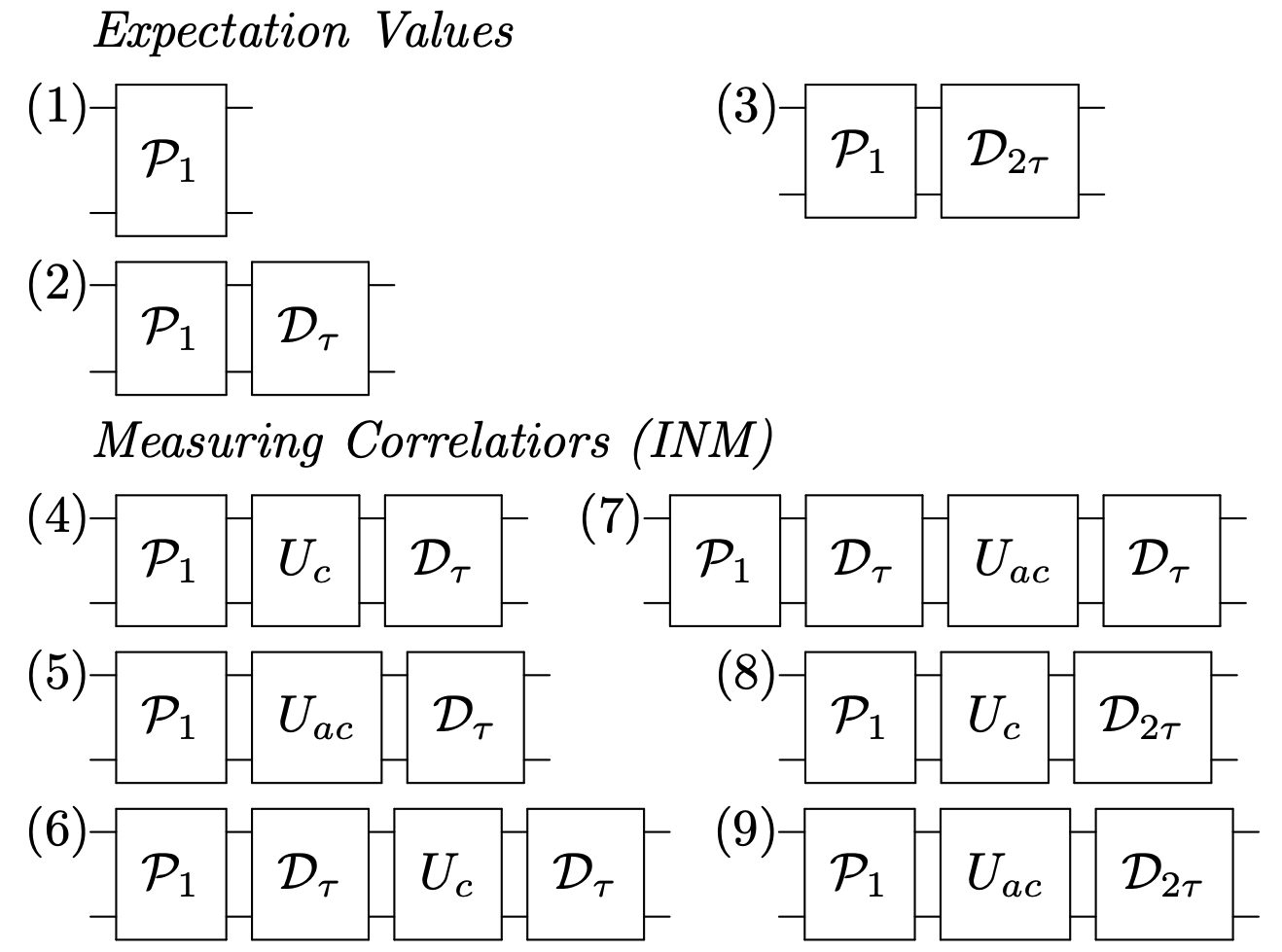}
    \caption{The first set of experiments. Experiments 1-3 are used to measure the values of $\langle Q_i \rangle$ and experiments 4-9 are used to measure the values of $C_{ij}$ using the INM protocol.}\label{fig:exp1INM}
\end{figure}

The first set of experiments (which look for violations of LG2 with LG3s satisfied) consists of nine pulse sequences which are depicted in in Fig.~ \ref{fig:exp1INM}. In experiments 1-3 the initial state is prepared and the system is evolved for different times before being measured. These measurements are used to determine $\langle Q_i \rangle$. The remaining experiments come in pairs,  4-5, 6-7 and 8-9. These experiments are used to determine the correlations. In each pair of experiments the measurement is done once with a CNOT gate and then once with an anti-CNOT gate. These measurements are done for the three different combinations of time intervals. For these experiments both the system and the ancilla are measured. This two qubit measurement is used to determine $C_{ij}$ as outlined in Section \ref{sec:INMTheory}.

\subsubsection{Second set of experiments}
\begin{figure}
    \includegraphics[width=\linewidth]{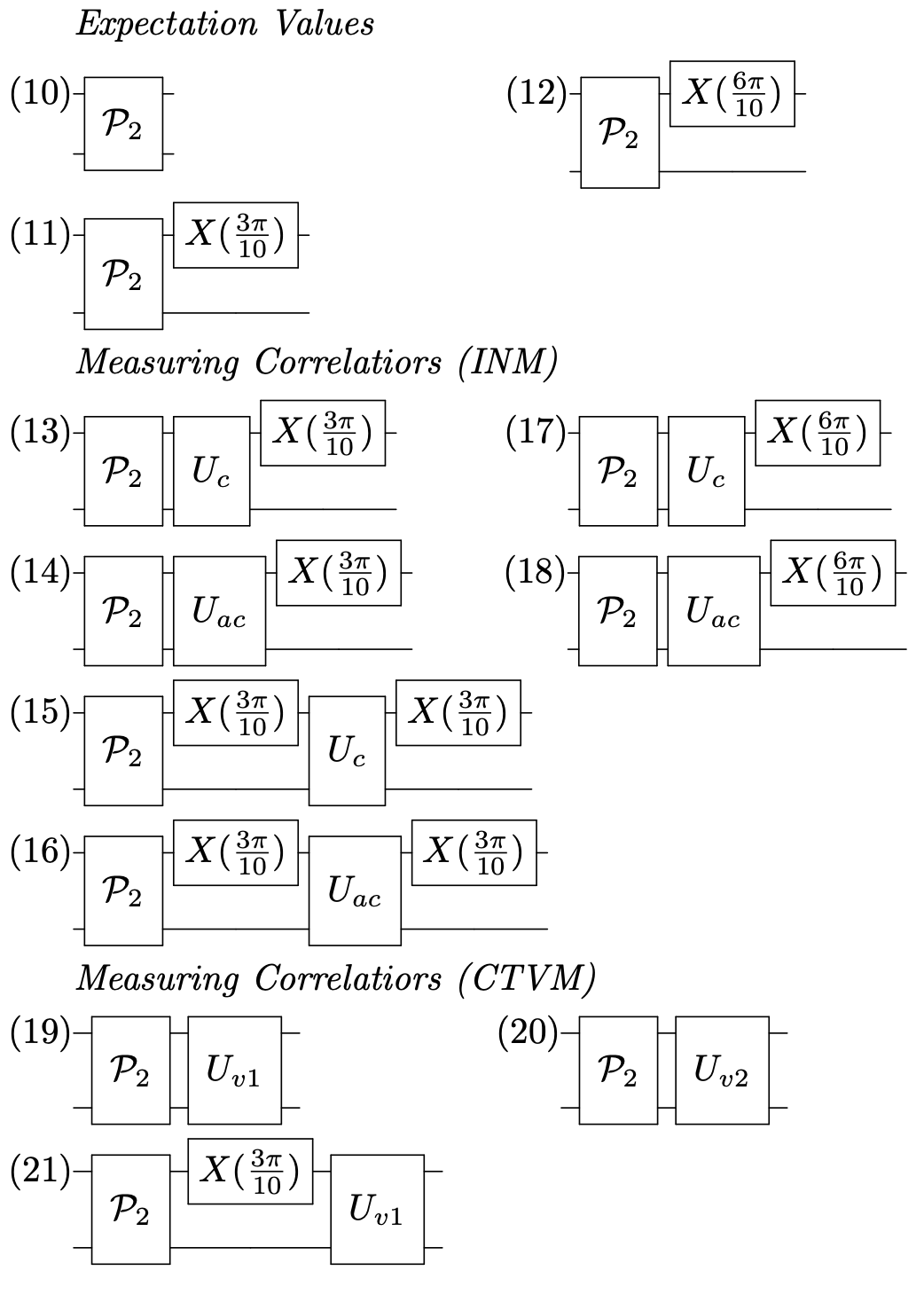}
    \label{fig:exp2CTVM}
    \caption{The second set of experiments. Experiments 10-12 are used to measure $\langle Q_i \rangle$, experiments 13-18 are used to measure the values of $C_{ij}$ using the INM protocol and experiments 19-21 are used to measure the values of $C_{ij}$ using the CTVM protocol.}
\end{figure}

The second set of experiments (which looks for violations of LG3 with LG2s satisfied) consists of twelve pulse sequences which are depicted in Fig.~\ref{fig:exp2CTVM}. In experiments 10-12 the initial state is prepared and the system is evolved for different times before being measured. These measurements are used to determine $\langle Q_i \rangle$ . Experiments 13-18 are used to determine the correlators $C_{ij}$. The logic for how these measurements are used to determine $C_{ij}$ follows exactly from the first set of  experiments. In experiments 19-21 the initial state is prepared and is then coupled to the ancilla through the system-detector component for different time intervals. The ancilla qubit is then measured to determine $C_{ij}$ as outlined in Section \ref{sec:CTVMTHEORY}.

\section{Experimental results} \label{sec:Experimental results}

\begin{table}
\def\arraystretch{1.25}
{\setlength{\tabcolsep}{0.65em}
\begin{tabular}{|c|c|c|c|}
\hline
\multirow{2}{*}{} & \multicolumn{1}{c|}{I} & \multicolumn{1}{c|}{N} & \multicolumn{1}{c|}{E}  \\  \hline
 \multicolumn{4}{|c|}{\textbf{First Set of Experiments}} \\ \cline{1-4} 
\multicolumn{4}{|c|}{Expectation values} \\ \cline{1-4} 
 $\langle Q_1 \rangle$ &  0.71  & 0.70  & 0.69 $\pm$ 0.02  \\
$\langle Q_2 \rangle$ &  0.45  & 0.45 & 0.43 $\pm$ 0.02  \\
$\langle Q_3 \rangle$ &  -0.61  & -0.60  & -0.58 $\pm$ 0.02 \\
\hline
\multicolumn{4}{|c|}{Correlators (INM)} \\ \cline{1-4} 
 $C_{12}$ & 0.00  & 0.00  & -0.01 $\pm$ 0.02  \\
$C_{23}$ & 0.00 & 0.00 & -0.01 $\pm$ 0.02 \\
$C_{13}$ &  -0.86  & -0.86 & -0.83 $\pm$ 0.02  \\
\hline 
 \multicolumn{4}{|c|}{\textbf{Second Set of Experiments}} \\ \hline
\multicolumn{4}{|c|}{Expectation values} \\ \cline{1-4} 
 $\langle Q_1 \rangle$ &  0.31  & 0.31 &  0.30 $\pm$ 0.02  \\
$\langle Q_2 \rangle$ & 0.18  & 0.18 & 0.18 $\pm$ 0.02  \\
$\langle Q_3 \rangle$ & -0.10  &  -0.09 & -0.06 $\pm$ 0.02  \\ \hline
\multicolumn{4}{|c|}{Correlators (INM)} \\ \hline
$C_{12}$ & 0.59  & 0.59  & 0.57 $\pm$ 0.03 \\
$C_{23}$ & 0.59  & 0.59 & 0.56 $\pm$ 0.02 \\
$C_{13}$ & -0.31  & -0.31  &  -0.29 $\pm$ 0.02  \\ \hline
\multicolumn{4}{|c|}{Correlators (CTVM)} \\ \hline
$C_{12}$ & 0.59  & 0.54  &  0.52 $\pm$ 0.07  \\
$C_{23}$ & 0.59  &  0.54 &  0.51 $\pm$ 0.08  \\
$C_{13}$ & -0.29  & -0.28  &  -0.24 $\pm$ 0.08  \\
 \hline
\end{tabular}}
\caption{The values of $\langle Q_i \rangle$ and $C_{ij}$ that are determined from the experimental data from the first and second set of experiments. (Ideal simulation (I), Noisy simulation (N) Experimentally determined (E)).}
\label{tbl:CijQi}
\end{table}

The experimental data from the first and second sets of experiments is provided in two tables in Appendix \ref{Experimental data}. The results from these two tables are used to calculate the values of $\langle Q_i \rangle$ and $C_{ij}$, which are provided in Table \ref{tbl:CijQi}. Table Table \ref{tbl:CijQi} also includes two simulated values for comparison. The first simulation, the ideal simulation, uses ideal pulses and assumes the existence of no natural decoherence. The second simulation, the noisy simulation, accounts for the added evolution from the systems natural Hamiltonian while pulses are being applied and also approximates the effect of the natural decoherence of the system. The procedure for determining the error bars on these values for both sets of experiments is outlined in Appendix \ref{App:Error}.

\begin{figure}
    \includegraphics[scale=0.5]{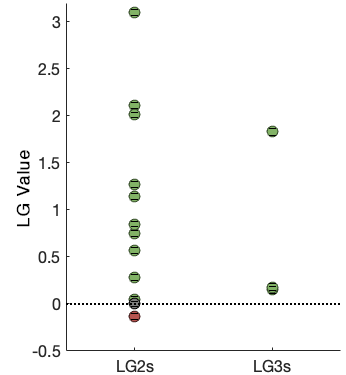}
    \caption{The values of the LG2s and LG3s constructed from the first set of experiments with their corresponding error bars. The labels in green highlight which inequalities were satisfied and the labels in red highlight which were violated.}
    \label{fig:Final1}
\end{figure}

\begin{figure}
    \includegraphics[scale=0.5]{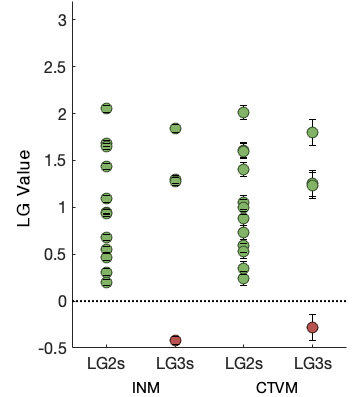}
    \caption{The values of the LG2s and LG3s constructed from the second set of experiments with their corresponding error bars. The labels in green highlight which inequalities were satisfied and the labels in red highlight which were violated.}
    \label{fig:Final2}
\end{figure}

The values of the $\langle  Q_i \rangle$ and $C_{ij}$ were used to determined the LG2s and LG3s for the two sets of experiments. The LG2s and LG3s for the first set of experiments are listed in Table \ref{tbl:Final1} and plotted in Fig.~\ref{fig:Final1}. Likewise, the LG2s and LG3s for the second set of experiments are listed in Table \ref{tbl:Final2} and plotted in Fig.~\ref{fig:Final2}.


\begin{table*}
\def\arraystretch{1.25}
{\setlength{\tabcolsep}{0.75em}
\begin{tabular}{|c|c|c|c|c|}
\hline
\multicolumn{1}{|c|}{\multirow{2}{*}{Label}} & \multirow{2}{*}{Inequality}  & \multicolumn{3}{c|}{INM} \\ \cline{3-5} 
\multicolumn{1}{|c|}{} &  & I& N & E \\
\hline \hline
{\color[HTML]{82B366} 2.1} &$1 + \langle  Q_1 \rangle  + \langle  Q_2 \rangle  + C_{12} \geq 0$ & 2.16 & 2.14 & 2.11 $\pm$ 0.03 \\
{\color[HTML]{82B366} 2.2}  &$1 - \langle  Q_1 \rangle + \langle  Q_2 \rangle - C_{12} \geq 0$ & 0.75 & 0.75 & 0.75 $\pm$ 0.03 \\
{\color[HTML]{82B366} 2.3}  &$1 + \langle  Q_1 \rangle - \langle  Q_2 \rangle - C_{12} \geq 0$ & 1.25 & 1.25 & 1.27 $\pm$ 0.03 \\
{\color[HTML]{B85450} \textbf{2.4}} &$1- \langle  Q_1 \rangle - \langle  Q_2 \rangle + C_{12} \geq 0$ & -0.16 & -0.14 & -0.13 $\pm$ 0.03 \\
{\color[HTML]{82B366} 2.5}  &$1 + \langle  Q_2 \rangle + \langle  Q_3 \rangle + C_{23} \geq 0$ & 0.85 & 0.85 & 0.85 $\pm$ 0.03 \\
{\color[HTML]{494949} 2.6}    &$1- \langle  Q_2 \rangle + \langle  Q_3 \rangle - C_{23} \geq 0$ & -0.06 & -0.05 & 0.00 $\pm$ 0.03 \\
{\color[HTML]{82B366} 2.7}  &$1 + \langle  Q_2 \rangle - \langle  Q_3 \rangle - C_{23} \geq 0$ & 2.06 & 2.05 & 2.01 $\pm$ 0.03 \\
{\color[HTML]{82B366} 2.8}  &$1- \langle  Q_2 \rangle - \langle  Q_3 \rangle + C_{23} \geq 0$ & 1.15 & 1.15 & 1.14 $\pm$ 0.03 \\
{\color[HTML]{82B366} 2.9}  &$1 + \langle  Q_1 \rangle + \langle  Q_3 \rangle + C_{13} \geq 0$ & 0.24 & 0.24 & 0.28 $\pm$ 0.03 \\
{\color[HTML]{82B366} 2.10} &$1- \langle  Q_1 \rangle + \langle  Q_3 \rangle - C_{13} \geq 0$ & 0.54 & 0.56 & 0.57 $\pm$ 0.03 \\
{\color[HTML]{82B366} 2.11} &$1 + \langle  Q_1 \rangle - \langle  Q_3 \rangle - C_{13} \geq 0$ & 3.17 & 3.16 & 3.10 $\pm$ 0.03 \\
{\color[HTML]{82B366} 2.12} &$1- \langle  Q_1 \rangle - \langle  Q_3 \rangle + C_{13} \geq 0$ & 0.04 & 0.04 & 0.05 $\pm$ 0.03 \\
\hline \hline
{\color[HTML]{82B366} 3.1} &$1 + C_{12} + C_{23} + C_{13} \geq 0$ & 0.14 & 0.15 & 0.15 $\pm$ 0.04 \\
{\color[HTML]{82B366} 3.2} &$1 - C_{12} - C_{23} + C_{13} \geq 0$ & 0.14 & 0.14 & 0.18 $\pm$ 0.04 \\
{\color[HTML]{82B366} 3.3} &$1 + C_{12} - C_{23} - C_{13} \geq 0$ & 1.86 & 1.86 & 1.83 $\pm$ 0.04 \\
{\color[HTML]{82B366} 3.4} &$1 - C_{12} + C_{23} - C_{13} \geq 0$ & 1.86 & 1.86 & 1.83 $\pm$ 0.04 \\
\hline
\end{tabular}}
\caption{The values of the LG2s and LG3s constructed from the first set of experiments. The labels in green highlight which inequalities were satisfied and the labels in red highlight which were violated (Ideal simulation (I), Noisy simulation (N) Experimentally determined (E)).}\label{tbl:Final1}
\end{table*}
\begin{table*}
\bgroup
\def\arraystretch{1.25}
{\setlength{\tabcolsep}{0.75em}
\begin{tabular}{|c|c|c|c|c|c|c|c|c|}
\hline
\multicolumn{1}{|c|}{\multirow{2}{*}{Label}} & \multirow{2}{*}{Inequality} &  \multicolumn{3}{c|}{INM} & \multicolumn{3}{c|}{CTVM} \\ \cline{3-8} 
\multicolumn{1}{|c|}{} &  & I &  N & E & I & N & E \\
\hline \hline
{\color[HTML]{82B366} 2.1} &$1 + \langle  Q_1 \rangle  + \langle  Q_2 \rangle  + C_{12} \geq 0$  & 2.08 & 2.07 & 2.05 $\pm$ 0.04 & 2.08 & 2.03 &  2.01 $\pm$ 0.08  \\
{\color[HTML]{82B366} 2.2}  &$1 - \langle  Q_1 \rangle + \langle  Q_2 \rangle - C_{12} \geq 0$ & 0.28 & 0.29 & 0.31 $\pm$  0.04 & 0.28 & 0.33 & 0.35 $\pm$ 0.08 \\
{\color[HTML]{82B366} 2.3}  &$1 + \langle  Q_1 \rangle - \langle  Q_2 \rangle - C_{12} \geq 0$ &0.54 & 0.54 & 0.55 $\pm$  0.04 &  0.54 & 0.58 & 0.59 $\pm$ 0.08 \\
{\color[HTML]{82B366} 2.4}  &$1- \langle  Q_1 \rangle - \langle  Q_2 \rangle + C_{12} \geq 0$ & 1.10 & 1.10 & 1.09 $\pm$  0.04 & 1.10 & 1.06  & 1.05 $\pm$ 0.08 \\
{\color[HTML]{82B366} 2.5}  &$1 + \langle  Q_2 \rangle + \langle  Q_3 \rangle + C_{23} \geq 0$ & 1.67 & 1.68 & 1.68 $\pm$ 0.03 & 1.68 & 1.63 & 1.61 $\pm$ 0.09 \\
{\color[HTML]{82B366} 2.6}  &$1- \langle  Q_2 \rangle + \langle  Q_3 \rangle - C_{23} \geq 0$ & 0.14 & 0.15 & 0.20 $\pm$ 0.03 & 0.13 & 0.19 & 0.24 $\pm$ 0.09 \\
{\color[HTML]{82B366} 2.7}  &$1 + \langle  Q_2 \rangle - \langle  Q_3 \rangle - C_{23} \geq 0$ & 0.69 & 0.68 & 0.68 $\pm$ 0.03 & 0.69 & 0.73  & 0.73 $\pm$ 0.09 \\
{\color[HTML]{82B366} 2.8}  &$1- \langle  Q_2 \rangle - \langle  Q_3 \rangle + C_{23} \geq 0$ &  1.50 & 1.49 & 1.44 $\pm$ 0.03 &1.50 & 1.45  & 1.40 $\pm$  0.09 \\
{\color[HTML]{82B366} 2.9}  &$1 + \langle  Q_1 \rangle + \langle  Q_3 \rangle + C_{13} \geq 0$ & 0.90 & 0.91 & 0.94 $\pm$ 0.03 & 0.92 & 0.94 & 1.00 $\pm$ 0.08 \\
{\color[HTML]{82B366} 2.10} &$1- \langle  Q_1 \rangle + \langle  Q_3 \rangle - C_{13} \geq 0$ & 0.90 & 0.91 & 0.93 $\pm$ 0.03 & 0.89 & 0.88  & 0.88 $\pm$ 0.08 \\
{\color[HTML]{82B366} 2.11} &$1 + \langle  Q_1 \rangle - \langle  Q_3 \rangle - C_{13} \geq 0$ & 1.71 & 1.70 & 1.65 $\pm$ 0.03 & 1.69 & 1.67  & 1.60 $\pm$ 0.08 \\
{\color[HTML]{82B366} 2.12} &$1- \langle  Q_1 \rangle - \langle  Q_3 \rangle + C_{13} \geq 0$ & 0.48 & 0.47 & 0.47 $\pm$ 0.03 & 0.50 & 0.51 & 0.53 $\pm$ 0.08 \\
\hline \hline
{\color[HTML]{82B366} 3.1} &$1 + C_{12} + C_{23} + C_{13} \geq 0$ & 1.87 & 1.86 & 1.84 $\pm$ 0.04 & 1.89 & 1.81   & 1.80 $\pm$ 0.14 \\
{\color[HTML]{B85450} \textbf{3.2}} &$1 - C_{12} - C_{23} + C_{13} \geq 0$ & -0.48 & -0.48  & -0.42 $\pm$ 0.04  &  -0.47 & -0.36 & -0.28 $\pm$ 0.14 \\
{\color[HTML]{82B366} 3.3} &$1 + C_{12} - C_{23} - C_{13} \geq 0$ & 1.31 & 1.31  & 1.30 $\pm$ 0.04  & 1.29 & 1.28  & 1.25 $\pm$ 0.14 \\
{\color[HTML]{82B366} 3.4} &$1 - C_{12} + C_{23} - C_{13} \geq 0$ &1.31 & 1.31 & 1.28 $\pm$ 0.04  &  1.29 & 1.28 & 1.23 $\pm$ 0.14 \\
\hline
\end{tabular}}
\egroup
\caption{The values of the LG2s and LG3s constructed from the second set of experiments. The labels in green highlight which inequalities were satisfied and the labels in red highlight which were violated (Ideal simulation (I), Noisy simulation (N) Experimentally determined (E)).}\label{tbl:Final2}
\end{table*}

The key points of the experimental results are as follows: As seen in Table \ref{tbl:Final1} and Fig.~\ref{fig:Final1}, in the first set of experiments the LG3s (labelled 3.1 - 3.4) were all satisfied and two of the LG2s (2.4 and 2.6) were violated. The violation of 2.4 is much more significant than that of 2.6 and since we only require one LG2 to be violated, we will focus on 2.4 as the violation of the LG2s. Furthermore, as seen in Table \ref{tbl:Final2}  and Fig.~\ref{fig:Final2}, for the seconds set of experiments the LG2s (labelled 2.1 - 2.12) were all satisfied and one of the LG3s (3.2) was violated. These two results experimentally demonstrate that neither the LG2s or LG3s are sufficient conditions for macrorealism. Additionally, for the second set of experiments both the CTVM and INM protocols were implemented and gave comparable results for the values of $C_{ij}$. These two protocols also provided the same violation and satisfaction of the corresponding LG2s and LG3s % The only point at which the values for the INM and CTVM protocols are not within one standard deviation of one another is for the LG3 violating point in Fig.~\ref{fig:Final2} (see (3.2) in Table \ref{tbl:Final2}).
%However, they both give very clear violations of the LG3, beyond experimental error,
and thus give the same qualitative conclusion. As calculated in Section \ref{Experiment_CTVM}, the LG3 violation for the CTVM protocol must violate the LGI bound by an extra value of 0.0028 for the violation to not be a result of the strength of the coupling constant, which it does.

\section{Conclusion}\label{sec:Conclusion}

The purpose of this work was twofold: to provide a more complete test of macroscopic realism using an augmented LG framework; to implement a new type of non-invasive measurement protocol and compare it to the standard ideal negative measurement protocol.
To date, LGI experiments have tested a set of conditions for MR formulated entirely in terms of temporal correlation functions at three pairs of times. These conditions for MR are necessary but not sufficient. The augmented LG inequalities considered here include an additional set of two-time inequalities which also involve the averages, $ \langle  Q_i \rangle$, and lead to a set of conditions which are both necessary and sufficient. In this work we showed how these conditions for MR could be tested experimentally. We exhibited experimentally situations in which the LG2s were satisfied but the LG3s violated, a natural parallel to the Bell case. We also exhibited situations in which the LG3s were satisfied but the LG2s violated, the key case in which the original LG framework based solely on LG3s fails to pick up violations of MR. Any future LG experiments should aspire to consider necessary and sufficient conditions for macrorealism, since a failure to violate the LG3s alone can not guarantee that a system can be described macrorealistically.
 
In this work we also performed the first implementation of the continuous in time velocity measurement protocol for determining correlation functions, a non-invasive technique very different to the usual ideal negative measurement protocol and with the advantage that it involves a different set of assumptions. First, it assumes that enough is known about the system to be able to identify the velocity corresponding to $Q$. Second, it assumes that the time intervals involved are sufficiently short so that, to a high probability, $Q$ will change sign only once. Lastly, it assumes that the coupling between the primary system and detector is sufficiently small that the back reaction of the detector on the future system dynamics is negligible. We argued that these three assumptions are easy to justify in the system we studied. In particular there is a regime in which they are satisfied in which there are also significant violations of the LG inequalities. Furthermore, we also found that in its domain of validity, the CTVM protocol agreed with the ideal negative measurement protocol.
 
A natural improvement of the CTVM protocol would be to use a detector with more than the two states used here. This would decrease the probability of error from multiple sign changes. This will be explored in future works.

We would like to point out that the main outcomes of this work are: (1) the first experimental implementation of a set of conditions for macrorealism using the LG2s and LG3s which, unlike earlier tests, are together necessary and sufficient conditions for macrorealism; (2) the experimental implementation of a novel approach to the non-invasive measurement of correlation functions formulated on a different set of assumptions to the traditional ideal negative measurement approach, and a confirmation that the two approaches agree. This work was not intended as an advance towards the goal of macroscopicity, but this clearly remains an important goal for future LG experiments.

\begin{acknowledgments}
We would like to thank Clive Emary, Joseph Emerson, George Knee, Johannes Kofler, Junan Lin, Owen Maroney, John Petterson, Rob Spekkens and James Yearsley for useful conversations. We would also like to thank Mike and Ophelia Lazaridis and the Canadian federal and Ontario provincial governments for funding this research.
\end{acknowledgments}

\appendix

\section{Measuring correlations from INM} \label{App:MeasuringCij}
In this section we demonstrate that the diagonal entries of the final output of the circuit in Fig.~\ref{fig:INMcirc} are precisely the values $aa', a(1\minus a'), (1\minus a)a''$ and $(1\minus a)(1\minus a'')$ that are required to determine the correlation functions.

We begin with an initial system that is evolving in time, coupled to the ancilla which is in the state $\ket{0}\bra{0}$. After evolving for some time $t_i$ this system is in the state
\begin{align}
\rho_1 & = U(t_i) \rho U(t_i)^{\dagger} \otimes \ket{0}\bra{0} \\
& = [\begin{smallmatrix} a & b \\ b^* & (1-a) \end{smallmatrix}] \otimes \ket{0}\bra{0}\\
&= \bigg[ \begin{smallmatrix} a&0&b&0 \\ 0&0&0&0 \\ b^*&0&1-a&0 \\ 0&0&0&0 \end{smallmatrix}\bigg]
\end{align}
the system then undergoes a CNOT
\begin{align}
\rho_2 &= \text{CNOT} \rho_1 \text{CNOT}^{\dagger}\\
& = \bigg[ \begin{smallmatrix} a&0&0&b \\ 0&0&0&0 \\ 0&0&0&0 \\ b^*&0&0&1-a   \end{smallmatrix}\bigg]
\end{align}
before evolving again for time $t_j - t_i$. Written most generally this evolution maps
$$ [\begin{smallmatrix} 1&0\\0&0 \end{smallmatrix}] \rightarrow [\begin{smallmatrix} a' & b' \\ b'^* & (1-a') \end{smallmatrix}], \quad [\begin{smallmatrix} 0&0\\0&1 \end{smallmatrix}] \rightarrow [\begin{smallmatrix} a'' & b'' \\ b''^* & (1-a'') \end{smallmatrix}]$$
The evolution of the states $\ket{0}\bra{1}$ and $\ket{1}\bra{0}$ will not effect the outcome of the diagonal elements, so we will drop these terms here. The evolution $U(t_j - t_i)$ will thus map the diagonal elements of $p_2$ to $\rho_3$ as
\begin{align}
\rho_2 &= [\begin{smallmatrix} 1&0\\0&0 \end{smallmatrix}] \otimes [\begin{smallmatrix} a&0\\0&0 \end{smallmatrix}] 
+ [\begin{smallmatrix} 0&0\\0&1 \end{smallmatrix}] \otimes [\begin{smallmatrix} 0&0\\0&(1-a) \end{smallmatrix}] 
\\
\rho_3 &= [\begin{smallmatrix} a' & b' \\ b'^* & (1-a') \end{smallmatrix}] \otimes [\begin{smallmatrix} a&0\\0&0 \end{smallmatrix}] 
+ [\begin{smallmatrix} a'' & b'' \\ b''^* & (1-a'') \end{smallmatrix}]\otimes [\begin{smallmatrix} 0&0\\0&(1-a) \end{smallmatrix}]
\end{align}
Which again, by considering just the diagonal terms becomes
\begin{equation}
\rho_3 = \Bigg[ \begin{smallmatrix} aa'&0&0&0 \\ 0&a(1-a')&0&0 \\ 0&0&(1-a)a''&0 \\ 0&0&0&(1-a)(1-a'')   \end{smallmatrix}\Bigg]
\end{equation}
Thus, the diagonal terms of the final state from Fig.~\ref{fig:INMcirc} are equivalent to the two-time probabilities required to determine the correlations $C_{ij}$.

\section{Theoretical result of INM} \label{App:TheoResultINM}
\begin{figure}
    \includegraphics[width=\linewidth]{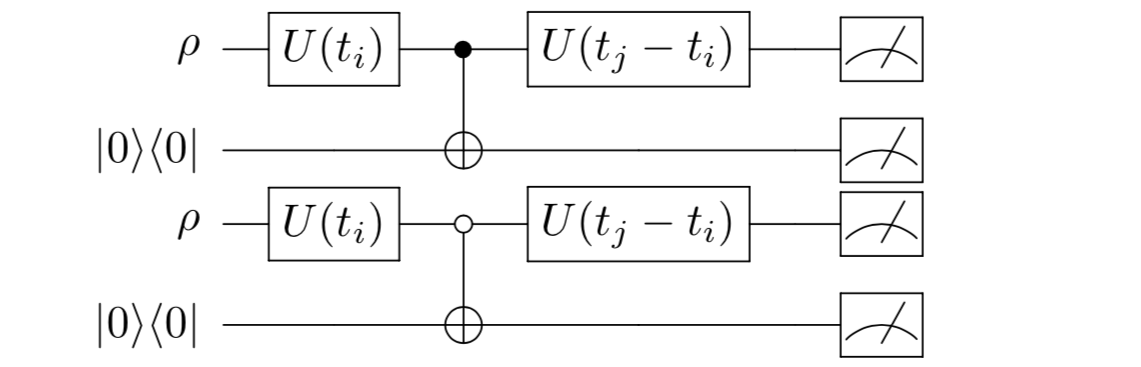}
    \caption{The two quantum circuits used to implement the INM procedure. }
    \label{fig:INMcirc2}
\end{figure}
In this appendix we demonstrate that the numerical result from the INM procedure outline in Section \ref{sec:INMTheory} match the ideal results of $C_{ij} = \cos(\omega (t_j - t_i))$ for the simple spin model.

We again consider our simple spin model, where the primary system evolves according to $H = \omega X/2$ and so $U(t_i) =e^{-i\omega Xt_i/2}$. The two circuits required to implement the INM protocol are given in Fig.~\ref{fig:INMcirc2}. We will first consider the CNOT circuit. At time $t_i$ the system exists in the general state
\begin{equation}
\rho(t_i) \otimes [\begin{smallmatrix} 1 & 0 \\ 0 &0 \end{smallmatrix}]= [\begin{smallmatrix} a & b\\b^* &(1-a) \end{smallmatrix}] \otimes [\begin{smallmatrix} 1 & 0 \\ 0 &0 \end{smallmatrix}]
\end{equation}
the initial state then undergoes the CNOT evolution to arrive at the state.
\begin{equation}
\rho_2= \text{CNOT}(\rho(t_i) \otimes [\begin{smallmatrix} 1 & 0 \\ 0 &0 \end{smallmatrix}])\text{CNOT}^{\dagger} =  \bigg[ \begin{smallmatrix} a&0&0&b \\ 0&0&0&0 \\ 0&0&0&0 \\ b^*&0&0&1-a   \end{smallmatrix}\bigg]
\end{equation}
the system then, from time $t_j - t_i \equiv t$, undergoes the evolution
\begin{align}
U(t) \otimes I &=e^{-i\omega Xt/2}  \otimes I \\
&= \cos \Big(\frac{\omega t}{2} \Big)II - i\sin\Big(\frac{\omega t}{2}\Big)XI
\end{align}
to arrive at the state $\rho_3$ ($c \equiv \cos \Big(\frac{\omega t}{2} \Big)$, $s \equiv \sin \Big(\frac{\omega t}{2} \Big)$)
\begin{align}
\rho_3 &= \bigg[ \begin{smallmatrix} c&0&\minus is&0 \\ 0&c&0&\minus is \\ \minus is&0&c&0 \\ 0&\minus is&0&c   \end{smallmatrix}\bigg] \bigg[ \begin{smallmatrix} a&0&0&b \\ 0&0&0&0 \\ 0&0&0&0 \\ b^*&0&0&1-a   \end{smallmatrix}\bigg] \bigg[ \begin{smallmatrix} c&0& is&0 \\ 0&c&0& is \\ is&0&c&0 \\ 0& is&0&c   \end{smallmatrix}\bigg] \\
& = \bigg[ \begin{smallmatrix}
ac^2&ibcs&iacs&bc^2 \\
-ib^*cs&-(a-1)s^2&b^*s^2&i(a-1)cs \\
-iacs&bs^2&as^2&-ibcs \\
b^*c^2&-i(a-1)cs&ib^*cs&-(a-1)c^2 \end{smallmatrix}\bigg] 
\end{align}
and by measuring the diagonal terms we find
\begin{align*}
p(+,+) = ac^2, \quad p(+,-) = (1-a)s^2 \\
p(-,+) = as^2, \quad p(-,-) = (1-a)c^2
\end{align*}
Since these values were determined using the CNOT circuit, to satisfy the INM protocol, only the $p(+,+)$ and $p(+,-)$ results can be kept. It is easy to show that these calculations can be repeated with the anti-CNOT circuit to find 
\begin{align}
\rho_3 &= \bigg[ \begin{smallmatrix} c&0&\minus is&0 \\ 0&c&0&\minus is \\ \minus is&0&c&0 \\ 0&\minus is&0&c   \end{smallmatrix}\bigg]
\bigg[ \begin{smallmatrix} 0&0&0&0 \\ 0&a&b&0 \\ 0&b^*&(1-a)&0 \\ 0&0&0&0   \end{smallmatrix}\bigg] \bigg[ \begin{smallmatrix} c&0& is&0 \\ 0&c&0& is \\ is&0&c&0 \\ 0& is&0&c   \end{smallmatrix}\bigg] \\
& = \bigg[ \begin{smallmatrix}
-(a-1)s^2&-ib^*cs&i(a-1)cs&b^*s^2 \\
ibcs&ac^2&bc^2&iacs \\
-i(a-1)cs&b^*c^2&-(a-1)c^2&ib^*cs \\
bs^2&-iacs&-ibcs&as^2 \end{smallmatrix}\bigg] 
\end{align}
and by measuring the diagonal terms we find
\begin{align*}
p(+,+) = ac^2, \quad p(+,-) = (1-a)s^2 \\
p(-,+) = as^2, \quad p(-,-) = (1-a)c^2
\end{align*}
Since these values were determined using the anti-CNOT circuit, to satisfy the INM protocol, only the $p(-,+)$ and $p(-,-)$ results can be kept. Thus the correlation function is
\begin{align}
C_{ij} & = p(++) - p(+-) - p(-+) + p(--)\\
& =  ac^2 -  (1-a)s^2 - as^2 + (1-a)c^2\\
& = c^2 - s^2\\
&= \cos \Big(\frac{\omega t}{2} \Big)^2 - \sin \Big(\frac{\omega t}{2} \Big)^2\\
& = \cos(\omega t) \\
&= \cos(\omega(t_j - t_i))
\end{align}
Thus providing the expected theoretical value for $C_{ij}$. 

\section{Experimental data} \label{Experimental data}

The experimental data from the first and second sets of experiments is provided in this appendix 

The  results of the measurements from the first sets of experiments are provided in Table \ref{tbl:data2}, and the results from the second set of experiments are provided in Table  \ref{tbl:data1}. These tables also provide the theoretically determined and simulated results for comparison. 

\begin{table}
\def\arraystretch{1.125}
{\setlength{\tabcolsep}{0.5em}
\begin{tabular}{|l|l|l|l|}
\hline
\multirow{2}{*}{} & \multicolumn{1}{c|}{T} & \multicolumn{1}{c|}{S} & \multicolumn{1}{c|}{E}  \\  \hline
\multicolumn{4}{|c|}{Expectation values} \\ \cline{1-4} 
\multicolumn{1}{|c|}{\multirow{2}{*}{(1)}}
& 0.85 & 0.85 & 0.84  $\pm$ 0.02 \\ \cline{2-4} 
& 0.15 & 0.15 & 0.16  $\pm$ 0.02 \\ \hline
\multicolumn{1}{|c|}{\multirow{2}{*}{(2)}}
& 0.73 & 0.72 & 0.71  $\pm$ 0.02 \\ \cline{2-4} 
& 0.27 & 0.28 & 0.29  $\pm$ 0.02 \\ \hline
\multicolumn{1}{|c|}{\multirow{2}{*}{(3)}}
& 0.20 & 0.20 & 0.21  $\pm$ 0.02 \\ \cline{2-4} 
& 0.80 & 0.80 & 0.79  $\pm$ 0.02 \\ \hline
\multicolumn{4}{|c|}{Correlators (INM)} \\ \cline{1-4} 
\multicolumn{1}{|c|}{\multirow{4}{*}{(4)}}
& 0.41 & 0.41 & 0.43  $\pm$ 0.02 \\ \cline{2-4} 
& 0.09 & 0.09 & 0.09  $\pm$ 0.02 \\ \cline{2-4} 
& 0.41 & 0.41 & 0.40  $\pm$ 0.02\\ \cline{2-4} 
& 0.09 & 0.08 & 0.09  $\pm$ 0.02 \\ \hline
\multicolumn{1}{|c|}{\multirow{4}{*}{(5)}}
& 0.09 & 0.09 & 0.08  $\pm$ 0.02 \\ \cline{2-4} 
& 0.41 & 0.41 & 0.39  $\pm$ 0.02 \\ \cline{2-4} 
& 0.09 & 0.09 & 0.09  $\pm$ 0.02\\ \cline{2-4} 
& 0.41 & 0.41 & 0.44  $\pm$ 0.02 \\ \hline
\multicolumn{1}{|c|}{\multirow{4}{*}{(6)}}
& 0.35 & 0.36 & 0.38  $\pm$ 0.02 \\ \cline{2-4} 
& 0.14 & 0.15 & 0.12  $\pm$ 0.02 \\ \cline{2-4} 
& 0.36 & 0.36 & 0.39  $\pm$ 0.02\\ \cline{2-4} 
& 0.14 & 0.14 & 0.11  $\pm$ 0.02 \\ \hline
\multicolumn{1}{|c|}{\multirow{4}{*}{(7)}} 
& 0.15 & 0.15 & 0.16  $\pm$ 0.02 \\ \cline{2-4} 
& 0.36 & 0.36 & 0.33  $\pm$ 0.02 \\ \cline{2-4} 
& 0.14 & 0.14 & 0.13  $\pm$ 0.02\\ \cline{2-4} 
& 0.36 & 0.35 & 0.38  $\pm$ 0.02 \\ \hline
\multicolumn{1}{|c|}{\multirow{4}{*}{(8)}}
& 0.04 & 0.04 & 0.02  $\pm$ 0.02 \\ \cline{2-4} 
& 0.16 & 0.16 & 0.13  $\pm$ 0.02 \\ \cline{2-4} 
& 0.77 & 0.77 & 0.78  $\pm$ 0.02\\ \cline{2-4} 
& 0.03 & 0.03 & 0.07  $\pm$ 0.02 \\ \hline
\multicolumn{1}{|c|}{\multirow{4}{*}{(9)}}
& 0.16 & 0.16 & 0.11  $\pm$ 0.02 \\ \cline{2-4} 
& 0.04 & 0.04 & 0.04  $\pm$ 0.02 \\ \cline{2-4} 
& 0.04 & 0.03 & 0.06  $\pm$ 0.02\\ \cline{2-4} 
& 0.77 & 0.77 & 0.79  $\pm$ 0.02 \\ \hline
\end{tabular}}
\caption{The results of the first set of experiments. The two diagonal elements of the systems qubit are recorded for experiments 1-3, while the four diagonal elements of the system and ancilla qubits are recorded for experiments 4-9. These values are used to determine  $\langle  Q_i \rangle$ and $C_{ij}$  as described in Section \ref{sec:Experimental results}. (Theoretical (T), Simulated (S) Experimentally determined (E))}
\label{tbl:data2}
\end{table}

\begin{table}
\def\arraystretch{1.07}
{\setlength{\tabcolsep}{0.5em}
\begin{tabular}{|l|l|l|l|}
\hline
\multirow{2}{*}{} & \multicolumn{1}{c|}{T} & \multicolumn{1}{c|}{S} & \multicolumn{1}{c|}{E}  \\  \hline
\multicolumn{4}{|c|}{Expectation values} \\ \cline{1-4} 
\multicolumn{1}{|c|}{\multirow{2}{*}{(10)}}
& 0.65 & 0.65 & 0.65  $\pm$ 0.02 \\ \cline{2-4} 
& 0.35 & 0.35 & 0.35  $\pm$ 0.02 \\ \hline
\multicolumn{1}{|c|}{\multirow{2}{*}{(11)}}
& 0.59 & 0.59 & 0.59  $\pm$ 0.02 \\ \cline{2-4} 
& 0.41 & 0.41 & 0.41  $\pm$ 0.02 \\ \hline
\multicolumn{1}{|c|}{\multirow{2}{*}{(12)}}
& 0.45 & 0.46 & 0.47  $\pm$ 0.02 \\ \cline{2-4} 
& 0.54 & 0.54 & 0.53  $\pm$ 0.02 \\ \hline
\multicolumn{4}{|c|}{Correlators (INM)} \\ \cline{1-4} 
\multicolumn{1}{|c|}{\multirow{4}{*}{(13)}}
& 0.52 & 0.52 & 0.48  $\pm$ 0.02 \\ \cline{2-4} 
& 0.07 & 0.07 & 0.09  $\pm$ 0.02 \\ \cline{2-4} 
& 0.13 & 0.14 & 0.18  $\pm$ 0.02\\ \cline{2-4} 
& 0.27 & 0.27 & 0.24  $\pm$ 0.02 \\ \hline
\multicolumn{1}{|c|}{\multirow{4}{*}{(14)}}
& 0.07 & 0.07 & 0.09  $\pm$ 0.02\\ \cline{2-4} 
& 0.52 & 0.52 & 0.48  $\pm$ 0.02 \\ \cline{2-4} 
& 0.27 & 0.27 & 0.30  $\pm$ 0.02 \\ \cline{2-4} 
& 0.13 & 0.14 & 0.13  $\pm$ 0.02 \\ \hline
\multicolumn{1}{|c|}{\multirow{4}{*}{(15)}} 
& 0.47 & 0.16 & 0.18  $\pm$ 0.02 \\ \cline{2-4} 
& 0.08 & 0.17 & 0.18  $\pm$ 0.02\\ \cline{2-4} 
& 0.12 & 0.04 & 0.05  $\pm$ 0.02\\ \cline{2-4} 
& 0.32 & 0.63 & 0.59  $\pm$ 0.02 \\ \hline
\multicolumn{1}{|c|}{\multirow{4}{*}{(16)}}
& 0.08 & 0.17 & 0.21  $\pm$ 0.02 \\ \cline{2-4} 
& 0.47 & 0.17 & 0.14  $\pm$ 0.02 \\ \cline{2-4} 
& 0.32 & 0.63 & 0.60  $\pm$ 0.02 \\ \cline{2-4} 
& 0.12 & 0.04 & 0.04  $\pm$ 0.02 \\ \hline
\multicolumn{1}{|c|}{\multirow{4}{*}{(17)}}
& 0.23 & 0.23 & 0.22  $\pm$ 0.02 \\ \cline{2-4} 
& 0.23 & 0.23 & 0.25  $\pm$ 0.02 \\ \cline{2-4} 
& 0.43 & 0.43 & 0.41  $\pm$ 0.02\\ \cline{2-4} 
& 0.12 & 0.12 & 0.12  $\pm$ 0.02 \\ \hline
\multicolumn{1}{|c|}{\multirow{4}{*}{(18)}} 
& 0.23 & 0.23 & 0.25  $\pm$ 0.02 \\ \cline{2-4} 
& 0.23 & 0.23 & 0.18  $\pm$ 0.02 \\ \cline{2-4} 
& 0.12 & 0.12 & 0.16  $\pm$ 0.02 \\ \cline{2-4} 
& 0.43 & 0.43 & 0.41  $\pm$ 0.02 \\ \hline
\multicolumn{4}{|c|}{Correlators (CTVM)} \\ \cline{1-4} 
\multicolumn{1}{|c|}{\multirow{2}{*}{(19)}}
& 0.990 & 0.989 & 0.988  $\pm$ 0.002 \\ \cline{2-4} 
& 0.010 & 0.011 & 0.012 $\pm$ 0.002 \\ \hline
\multicolumn{1}{|c|}{\multirow{2}{*}{(20)}}
& 0.990 & 0.989 & 0.988  $\pm$ 0.002 \\ \cline{2-4} 
& 0.010 & 0.011 & 0.012 $\pm$ 0.002 \\ \hline
\multicolumn{1}{|c|}{\multirow{2}{*}{(21)}}
& 0.969 & 0.969 & 0.970  $\pm$ 0.002 \\ \cline{2-4} 
& 0.031 & 0.031 & 0.030  $\pm$ 0.002 \\ \hline
\end{tabular}}
\caption{The results of the second set of experiments. The two diagonal elements of the systems qubit are recorded for experiments 10-12, the four diagonal elements of the system and ancilla qubits are recorded for experiments 13-18 and the two diagonal elements of the ancilla qubit are recorded for experiments 19-21. The values from experiments 10-12 are used to determine  $\langle  Q_i \rangle$, the values from experiments 13-18 are used to determine $C_{ij}$ with the INM protocol and the values from experiments 19-21 are used to determine $C_{ij}$ with the CTVM protocol as described in Section \ref{sec:Experimental results}. (Theoretical (T), Simulated (S) Experimentally determined (E)) system system and ancilla qubits are recorded for experiments 13-18. These val}
\label{tbl:data1}
\end{table}

For the results of the first set of experiments either two or four measured values are provided. This is because experiments (1-3) were used for determining $\langle  Q_i \rangle$ and this only requires measuring the diagonal values of the first qubits density matrix. This can be seen by considering the measurement of $\langle  Q_i \rangle$ for a general state $\rho(t_i) = [\begin{smallmatrix} a & b\\ b^* & (1-a) \end{smallmatrix}]$ 
\begin{align}
\langle  Q_i \rangle &= \text{tr}(Z \rho(t_i)) \\
&=   \text{tr}(Z [\begin{smallmatrix} a & b\\ b^* & (1-a) \end{smallmatrix}])\\
& = 2a - 1 \label{eqn:Qdepends}
\end{align}

Experiments (6-9) required measuring the diagonal values of both the system and the ancilla. This is because, as shown in Section \ref{sec:INMTheory}, only these values are necessary to determine $C_{ij}$. Experiments (1-3) were used for determining $C_{ij}$. This require, as shown in Section \ref{sec:INMTheory}, only measuring the diagonal values of the density matrix of the system and ancilla.

Similarly, only two or four measured values are provided for the second set of experiments as well. Since experiments (10-18) also use the INM protocol, the logic for providing two or four values follows exactly as from experiments (1-9). On the other hand, Experiments (19-24) use the CTVM protocol. Of these experiments, numbers (19-21) are used to determining $\langle  Q_i \rangle$ in identical fashion to the INM protocol, and as such again only require two values. Unlike the INM protocol, the CTVM protocol again only requires two values to determine $C_{ij}$. As shown from Eq.(\ref{eqn:cijp(1)}) the value of $C_{ij}$ for some general state $\rho(t_i) = [\begin{smallmatrix} a & b\\ b^* & (1-a) \end{smallmatrix}]$ can be determined from only the diagonal entries of the ancilla qubit.
\begin{align}
C_{ij} &= 1 - \frac{p(1)}{2\lambda^2}\\
&= 1 - \frac{(1-a)}{2\lambda^2}
\end{align}
Thus, only two values are measured for experiments (22-24). The procedure for determining the error bars on these values for both sets of experiments is outlined in Appendix \ref{App:Error}.

\section{Error analysis} \label{App:Error}

The error bars on the experimentally determined values for the experiments performed in this work were determined by considering two potential sources of error, the first source being from the calibration of the pulses. Note that the degree of rotation from an NMR pulse depends on the power of the pulse and the length of its duration. The calibration procedure consisted of running forty experiments that each consisted of implementing two $Y$ pulses on the thermal state.  Each experiment used the same pulse power but varying pulse durations, $\tau$, starting from $\tau = 0 \mu s$. Furthermore, we note that the thermal state produces no signal on its own, produces a maximum signal when rotated $\frac{\pi}{2}$ radians and produces again no signal when rotated $\pi$ radians. Thus, as $\tau$ increases the signal reaches a maximum value before decreasing again to zero. The strength of the signal was plotted as a function of $\tau$ and then fitted to the function $f(\tau) = a \cos(b\tau) + c$ to find the $\tau$ in which a $\pi$ rotation occurred. Since two pulses were implemented, this procedure provides a value of $\tau$ corresponding to a $\frac{\pi}{2}$ rotation. The difference between the experimental data and $f(\tau)$ was taken for each of the forty experiments. These values were squared, summed and then square rooted to find the potential error from calibration. The second source of error that was considered was from the noise of the signal. A band of data points which should ideally produce no signal were assessed. The square root of the sum of the squares of the data points in the band was calculated and divided by the total area of the signal to provide a percent error. This percent error was also incorporated into the error bars. The final source of error which was considered in this work was from the natural drift of the NMR spectrometers magnetic field through out consecutive experiments. Since the spectrometers magnetic field gradually drifts, periodic breaks are taken between experiments to manually realign the magnetic field, a process known as \textit{shimming}. To account for potential errors introduced from this the spectra for the pps state was taken between experiments to see the range of possible values the pps would take. This range of values was taken into account in the error bars on the data.

\nocite{*}

\bibliography{apssamp}% Produces the bibliography via BibTeX.

\end{document}